\documentclass[12pt]{article}
	\pdfoutput=1 

\usepackage{amsmath}
\usepackage{amssymb}
\usepackage{amsfonts}
\usepackage{graphicx}
\usepackage[utf8]{inputenc}			
\usepackage{aas_macros}				
\usepackage{bm}                 

\usepackage{slashed}				
\usepackage{mathrsfs}				
\usepackage{bbm}					
\usepackage{cancel}					

\usepackage[dvipsnames]{xcolor}
\usepackage[hang,flushmargin]{footmisc} 

\usepackage{fancyhdr}		
\usepackage{lipsum}			
\usepackage{framed}			
\usepackage{subcaption}		
\usepackage{cite}			
\usepackage{xspace}			

\usepackage{booktabs}		

\usepackage[font=small]{caption} 
\usepackage{float}         

\usepackage[margin=2cm]{geometry}   
\graphicspath{{figures/}}			
\numberwithin{equation}{section}    

\usepackage{multicol}
\usepackage{etoolbox}
\usepackage{relsize}
\patchcmd{\thebibliography}
  {\list}
  {\begin{multicols}{2}\smaller\list}
  {}
  {}
\appto{\endthebibliography}{\end{multicols}}

\let\oldenumerate\enumerate
\renewcommand{\enumerate}{
  \oldenumerate
  \setlength{\itemsep}{3pt}
  \setlength{\parskip}{0pt}
  \setlength{\parsep}{0pt}
}

\let\olditemize\itemize
\renewcommand{\itemize}{
  \olditemize
  \setlength{\itemsep}{3pt}
  \setlength{\parskip}{0pt}
  \setlength{\parsep}{0pt}
}

\usepackage{lineno}

\renewcommand{\tilde}{\widetilde}   
\newcommand{\UV}{\textnormal{UV}}
\newcommand{\IR}{\textnormal{IR}}

\newcommand{\email}[1]{\href{mailto:#1}{#1}}
\newenvironment{institutions}[1][2em]{\begin{list}{}{\setlength\leftmargin{#1}\setlength\rightmargin{#1}}\item[]}{\end{list}}

\usepackage[
	colorlinks=true,
	citecolor=green!50!black,
	linkcolor=NavyBlue!75!black,
	urlcolor=green!50!black,
	hypertexnames=false]{hyperref}

\fancypagestyle{firststyle}{
	\rhead{\footnotesize%
	\texttt{UCR-TR-2020-FLIP-IG-11}%
	}}

\usepackage{etoc}

\newcommand{\pmd}{p_1}
\newcommand{\pnd}{p_2}

\usepackage{changes}

\begin{document}

	\thispagestyle{empty}		
	\thispagestyle{firststyle} 	

\begin{center}
    
    \phantom{.}
    \vspace{1.5cm}

    \textbf{\Large
    Effective Field Theory in AdS:
    \\[.2em]Continuum Regime, Soft Bombs, and IR Emergence 
    }

    \vskip .7cm

   { \bf 
   	Alexandria~Costantino$^{a}$,
   	Sylvain~Fichet$^{b, c}$,
   	and 
   	Philip~Tanedo$^{a}$ 
   	} 
   \\ 
   \vspace{-.2em}
   { \tt \footnotesize
	    \email{acost007@ucr.edu},
	    \email{sfichet@caltech.edu}, 
	    \email{flip.tanedo@ucr.edu}
   }
	

   \begin{institutions}[2.25cm]
   \footnotesize
   $^{a}$ 
   {\it 
	    Department of Physics \& Astronomy, 
	    University of  California, Riverside, 
	    {CA} 92521
	    }    
	\\ 
	\vspace*{0.05cm}   
	$^{b}$ 
	{\it 
       Walter Burke Institute for Theoretical Physics, 
       California Institute of Technology,\\
       \quad \hspace{-.3em}Pasadena, CA 91125
       }
     \\ 
	\vspace*{0.05cm}   
	$^{c}$ 
	{\it 
       ICTP South American Institute for Fundamental Research \& IFT-UNESP,\\
       \quad \hspace{-.3em}R.~Dr.~Bento Teobaldo Ferraz 271, S\~ao Paulo, Brazil
       }
   \end{institutions}

\end{center}
\vspace{.5cm}


\begin{abstract}
\noindent
We consider a scalar field in a slice of Lorentzian five-dimensional AdS at arbitrary energies.  
We show that the presence of bulk interactions separate the behavior of the theory into two different regimes:  Kaluza--Klein and continuum.
We determine the transition scale between these regimes and show that UV brane correlation functions are independent of IR brane-localized operators for four-momenta beyond this transition scale.
The same bulk interactions that induce the transition also give rise to cascade decays.
We study these cascade decays for the case of a cubic self-interaction in the continuum regime. 
We find that the cascade decay progresses slowly towards the IR region and gives rise to soft spherical final states,  in accordance with former results from both gravity and CFT. 
We identify a recursion relation between integrated squared amplitudes of different leg numbers and thus evaluate the total rate.
We find that cascade decays in the continuum regime  are  exponentially suppressed.
This feature completes the picture of the IR brane as an emergent sector as seen from the UV brane. We briefly discuss consistency with the holographic dual description of glueballs and some implications for dark sector models. 
\end{abstract}

\clearpage
\small
\setcounter{tocdepth}{2}
\tableofcontents
\normalsize
\clearpage


\section{Introduction }
\label{se:intro}

The AdS/CFT correspondence states that gauge theories with  large 't~Hooft coupling are dual to a weakly coupled string theory with a curved extra dimension~\cite{Maldacena:1997re,Gubser:1998bc,Witten:1998qj,Aharony:1999ti}. For sufficiently large 't~Hooft coupling, string states are heavy and the 5D theory is described by an effective field theory living in an AdS$_5$ background, see {e.g.}~\cite{Heemskerk:2009pn,Heemskerk:2010ty,Fitzpatrick:2010zm,Sundrum:2011ic,ElShowk:2011ag,Fitzpatrick:2012cg}. 
Variations of this duality can deform and even truncate the IR region of the AdS space, leading to a discrete tower of Kaluza--Klein states analogous to the hadron spectrum in QCD~\cite{Karch:2006pv,Gursoy:2007cb,Gursoy:2007er,Gubser:2008ny,Falkowski:2008fz, Batell:2008zm, Batell:2008me,Cabrer:2009we,vonGersdorff:2010ht, Cabrer:2011fb,Megias:2019vdb}).
Unlike QCD---which has small 't~Hooft coupling---cascades of radiation at large 't\,Hooft coupling do not form jets because there is no reason for soft or collinear phase space configurations to be preferred. Instead, cascades produce spherical events ending in a large number of low-momentum final states~~\cite{Hofman:2008ar, Chesler:2008wd, Hatta:2007cs,Hatta:2008tx, Hatta:2008qx,Hatta:2008tn, Knapen:2016hky}. 
Earlier field-theoretical studies of these \emph{soft bomb} events in AdS$_5$ assume that Kaluza--Klein modes are narrow~\cite{Csaki:2008dt}. 
In this work we show that around some transition scale, the narrow modes merge to form a continuum. 
We extend the study of soft bombs into this \emph{continuum regime}.

The regimes of field theory in a slice of 5D AdS, as obtained in this work, are the following.
The fundamental scales fixed by geometry are the AdS curvature, $k$, and the IR brane position, $1/\mu$.
The Kaluza--Klein scale is $\mu \ll k$ and represents the mass gap in the dual gauge theory. In the presence of interactions, the theory has a  5D cutoff $\Lambda$ and a transition scale $\tilde \Lambda$ that is explained below. These have a hierarchy $\Lambda>k>\tilde \Lambda> \mu$ that define four different energy regimes:
\begin{itemize}
    \item {4D regime},  $E<\mu$. In this limit, Kaluza--Klein modes are integrated out and only sufficiently light 4D modes such as gauge or Goldstone bosons remain in the spectrum.
    \item {Kaluza--Klein regime},  $\mu< E < \tilde \Lambda$. The theory in this regime has a tower of regularly spaced narrow resonances.  
    {The resonances in this energy window are narrow glueballs in the dual gauge theory.}
    \item {Continuum regime},
    $ \tilde \Lambda < E < k$. In this regime, the effective theory breaks down in the IR region of AdS. Quantum corrections mix the KK modes and merge them into a continuum. An observer on the UV brane effectively sees pure AdS. 
    {The theory can equivalently be described by a holographic CFT with no mass gap.}
    \item Flat space regime, $k < E < \Lambda$.
    Here the curvature of AdS becomes negligible, and KK modes  from any other compact dimensions   appear.
    {No simple CFT dual is expected in this regime.}
    
\end{itemize}

The overarching idea throughout this work is that the bulk correlators effectively lose contact with the IR brane in the continuum regime. 
At the qualitative level, this can be argued from the breakdown of the effective theory in the IR region (using e.g. Refs. \cite{ArkaniHamed:2000ds,Goldberger:2002cz}). 
Quantitatively, this requires one to account for bulk loops that dress the propagator~\cite{Fichet:2019hkg}.
Throughout this work we say that the IR brane \textit{effectively emerges} for bulk correlators as their energy is decreased through this KK--continuum transition.

One may conjecture that for $p\gg \tilde \Lambda$, the theory is described by an effective Lagrangian without an IR brane. A workaround to this conjecture may be possible via bulk cascade diagrams (i.e.\ soft bombs). A cascade diagram can split the energy of an individual state in the continuum regime into many offspring states reaching into the KK regime.
The evaluation of the rate for such event would imply that the high energy theory already knows about the IR brane,  which would invalidate the proposal that there is effectively no IR brane in the continuum regime. A careful investigation of the soft bomb rates is thus required.

In summary, this work \textit{i)} establishes the existence of a continuum regime in the presence of interactions and \textit{ii)} studies soft bomb events in this regime.
There are multiple motivations for such a study:
\begin{itemize}
    \item The earlier work on soft bombs in the Kaluza--Klein regime~\cite{Csaki:2008dt} does not apply in the continuum regime because the effective theory breaks down in the IR region of AdS. KK modes are thus not appropriate degrees of freedom.
    We thus investigate whether events are indeed spherical and soft in the continuum regime. This also serves as a check of the soft bomb picture in the CFT dual.

    \item In addition to determining  kinematic features, we seek to calculate occurrence probabilities for soft bomb events. To the best of our knowledge, such a calculation has not been presented in the literature.
    
    \item Understanding the KK--continuum transition and the soft bomb rate allows us to complete the picture of the emergence of the IR brane. Without {knowledge of the} soft bomb rates, it remains unclear whether the theory can actually be described by a high-energy effective theory with no IR brane in the continuum regime. 
    
    \item Both IR brane emergence and the properties of soft bombs have phenomenological implications for models of physics beyond the Standard Model that involve a strongly--coupled hidden sector with an AdS dual. This holographic dark sector scenario has been recently put presented in \cite{Brax:2019koq,Costantino:2019ixl}, see also~\cite{vonHarling:2012sz,McDonald:2012nc,McDonald:2010fe,McDonald:2010iq, Katz:2015zba,Strassler:2008bv} for earlier and related attempts. 
\end{itemize}

This paper is organized as follows.
Section~\ref{se:AdS} establishes the basic five-dimensional formalism in a slice of AdS. In particular, we present the classical propagator for a scalar field in mixed position--momentum space.
Interactions in the bulk of AdS play a central role in our study. Section~\ref{se:NDA} provides the necessary tools for dimensional analysis at strong coupling. 
In  Section~\ref{se:regimes}, we dress the propagator with quantum corrections.
The imaginary part of the self-energy induces distinct KK and continuum regimes. The transition scale is understood both qualitatively from the viewpoint of effective theory validity and from the viewpoint of the opacity of the IR region resulting from the dressing of the propagator by bulk fields. 
In Section~~\ref{se:Decays} we identify a recursion relation that relates the continuum-regime cascade decay rates with arbitrary number of legs. 
Section~\ref{se:picture} presents the general picture of soft bomb events in the continuum regime. 
Building on this, we spell out the notion of IR brane emergence. Asymptotically AdS backgrounds and implications for holographic dark sectors are also discussed. 
In Section \ref{se:adscft} we  connect our analysis to strongly coupled gauge theories using AdS/CFT.  We discuss CFT soft bombs,
establish the relation between bulk matter interactions and large-$N$ expansion, and analyse the transition scale in the EFT  of glueballs.
Conclusions are given in Sec.\,\ref{se:con}.

\section{A Bulk Scalar in a Slice of AdS}
\label{se:AdS}

In studies of the gravity--scalar system, a general ansatz for the metric  preserving  the 4D Poincar\'e invariance is 
\begin{align}
	ds^2	=g_{MN} \, dX^{M}dX^{N}
			=e^{-2A(y)} \eta_{\mu\nu}dx^\mu dx^\nu-dy^2   \ ,
\label{eq:metric:general}
\end{align}
where $\eta_{\mu\nu}$ is the 3+1-dimensional Minkowski metric with $(+,-,-,-)$ signature.
This metric appears in certain 5D supergravities, see \textit{e.g.} \cite{Freedman:1999gp}. 
It  can depart from AdS and develop a singularity at large $y$, beyond which spacetime ends, see \textit{e.g.} \cite{Karch:2006pv,Gursoy:2007cb,Gursoy:2007er, Batell:2008zm, Cabrer:2009we,vonGersdorff:2010ht}.  In other classes of models, an IR brane truncating the $y$ coordinate is explicitly included. 
In this paper we focus on the simplest example of a slice of AdS for which the metric is exactly anti-de Sitter.
Using the conformally flat coordinates $z=e^{ky}/k$, the metric is 
\begin{align}
	ds^2	=g_{MN} \, dX^{M}dX^{N}
			=(kz)^{-2}\left( \eta_{\mu\nu}dx^\mu dx^\nu-dz^2 \right)  \ .
\label{eq:metric}
\end{align} 
Space is truncated  at endpoints 
\begin{align}
	z_\textnormal{UV} &= {k^{-1}}
	& \textnormal{and}&&
	z_\textnormal{IR} &= {\mu^{-1}} > z_\textnormal{UV} \ ,
\end{align}
which correspond to the positions of a UV and IR brane, respectively.

\subsection{Action}

A generic effective theory on this background involves gravitons and matter fields of different spins. In this manuscript we focus on the case of a scalar field $\Phi$ with non-derivative, cubic interactions. We expect that the results of this study generalize readily to any other type of field.
The action for this field is
\begin{align}
	S &= 
	\int d^5X \sqrt{g}
	\left(
		\frac{1}{2} \nabla_M\Phi \nabla^M\Phi 
		- \frac{1}{2} m^2_\Phi \Phi^2 
		+ \frac{1}{3!} \lambda\Phi^3
	\right)
	+ S_{\rm UV}
	+ S_{\rm IR}
	+ \cdots \, \label{eq:SPhi}
\end{align}
where we explicitly write the kinetic, mass and interaction terms. 
The ellipses denote additional contributions from gravity and higher-dimensional operators that are suppressed by powers of the effective theory's cutoff. 
A convenient parameterization of the scalar mass is
\begin{align}
	m_\Phi^2 \equiv (\alpha^2-4)k^2 \ .
\end{align}
The Breitenlohner-Freedman bound requires $\alpha^2\geq 0$~\cite{Breitenlohner:1982jf,Breitenlohner:1982bm}. In this work we routinely take $\alpha$ to be non-integer.
The actions $S_\textnormal{UV}$ and $S_\textnormal{IR}$ encode brane-localized operators. 
These can include mass terms for the scalar which are conveniently parameterized with respect to dimensionless parameters $b_\UV$ and $b_\IR$ as~(see, e.g.~\cite{Ponton:2012bi}),
\begin{align}
	S_{\UV} + S_{\IR} 
	\supset 
	\frac{1}{2} 
	\int d^5X \sqrt{\bar{g}} 
	\left[
		(\alpha-2-b_{\UV})k\delta(z-z_\UV)-(\alpha-2+b_{\IR})k\delta(z-z_\IR)
	\right]
	\Phi^2
	\, .
\end{align}
We leave these parameters unspecified and simply assume that $b_\UV \neq 0$. There is a special mode in the spectrum with mass $\sim b_{\text{UV}} k$. For $b_{\text{UV}}$ sufficiently small, this mode may affect the physical processes studied here. We assume this special mode is heavy such that it is irrelevant in our analysis.
$\bar{g}_{\mu\nu}$ is the induced metric on the brane so that $\sqrt{\bar{g}}=(kz)^{-4}$. 
Other degrees of freedom may be localized on the brane and interact with $\Phi$.~\footnote{There are hints that a brane-localized degree of freedom always arises from a bulk field and is thus necessarily accompanied by a tower of Kaluza-Klein modes~\cite{Fichet:2019owx}. This tower can be decoupled from the brane so that it is consistent to consider only the brane-localized mode. }  In the context of our analysis, such brane modes provide asymptotic states for the bulk scattering amplitudes.

\subsection{The Scalar Propagator \label{se:propa}}

The classical equation of motion obtained by varying the bulk action for the scalar field, $\Phi$, is
\begin{align}D\Phi\equiv\frac{1}{\sqrt{g}}\partial_M(g^{MN}\sqrt{g}\partial_N\Phi)+m^2_\Phi\Phi=0 \ .
\end{align}
The Feynman propagator is the Green's function of the $D$ operator,
\begin{align}
	D_X\Delta(X,X')
	= \frac{-i}{\sqrt{g}} \delta^{(5)}(X-X') \ . 
	\label{eq:DDelta_free}
\end{align}
{Rather than work in position-space coordinates, $X^M = (x^\mu,z)$, we Fourier transform along the 4D Minkowski slices: $\Phi_p(z)\equiv \int e^{i \eta_{\mu\nu}x^\mu p^\nu} \Phi(x^\mu, z) $. 
We call this Poincar\'e position-momentum space. The AdS dilatation isometry becomes $(p^\mu, z)\rightarrow (p^\mu /\lambda, \lambda z)$ so that $pz$ is an invariant. Here $p$ is the Minkowski norm $p=\sqrt{\eta_{\mu\nu}p^\mu p^\nu}$, which is real (imaginary) for timelike (spacelike) four-momentum, $p^\mu$. 
In these coordinates, the propagator is, see \textit{e.g. }\cite{Fichet:2019owx},
}
\begin{align}
	\Delta_p(z,z')=&
 	i \frac{\pi k^3 (zz')^2}{2}
 	\frac{
 		\left[\tilde{Y}^{UV}_\alpha J_\alpha(pz_<)-\tilde{J}^{UV}_\alpha Y_\alpha(pz_<)\right]\left[\tilde{Y}^{IR}_\alpha J_\alpha(pz_>)-\tilde{J}^{IR}_\alpha Y_\alpha(pz_>)\right]
 	}{
 		\tilde{J}^{UV}_\alpha\tilde{Y}^{IR}_\alpha-\tilde{Y}^{UV}_\alpha\tilde{J}^{IR}_\alpha
 	}\ , 
 \label{eq:propa}
 \end{align} 
where
$z_{<,>}$ is the lesser/greater of the endpoints $z$ and $z'$. 
The $p$-dependent quantities $\tilde{J}^{UV,IR}$ are
\begin{align}
\tilde{J}^{UV}_\alpha = \frac{p}{k} J_{\alpha-1}\left(\frac{p}{k}\right) - b_{UV}J_\alpha\left(\frac{p}{k}\right)&\quad&
\tilde{J}^{IR}_\alpha = \frac{p}{\mu} J_{\alpha-1}\left(\frac{p}{\mu}\right) + b_{IR}J_\alpha\left(\frac{p}{\mu}\right),
\end{align}
with similar definitions for $\tilde{Y}^{UV,IR}$.

For timelike momentum, the propagator \eqref{eq:propa} has poles set by the zeros of the denominator. 
This propagator can always be written formally as an infinite sum over 4D poles. Let us introduce the matrix notation 
\begin{align}
	\mathbf{f}(z) &=
	\left[\;
		f_n(z)
	\;\right]
	&
	\mathbf{D}&=
	\left[\;
		\frac{ \delta_{nr}}{p^2-m_n^2}
	\;\right]
	\ ,
\end{align}
where ${\bf f}$ is a one-dimensional infinite vector and ${\bf D}$ is an infinite diagonal matrix indexed by the Kaluza--Klein (KK) numbers $n$ and $r$.
The propagator in the Kaluza-Klein representation is
\begin{align}
\Delta_p(z,z')=i\,{\bf f}(z)\cdot {\bf D} \cdot  {\bf f}(z')\,.
\end{align}

Amplitude calculations often feature sums over KK modes. We can represent these sums as contour integrals~\cite{Fichet:2019hkg}, 
\begin{align}
	\sum_{n=0}^{\tilde n}U(m_n)f_n(z)f_n(z')
	=
	-\frac{1}{2\pi} 
	\oint_{\mathcal C \left[\tilde n\right]}
	dq^2 U(q^2) \Delta_q(z,z')
	\ ,
	\label{eq:sumtoint}
\end{align}
where the contour $\mathcal C \left[\tilde n\right]$ in momentum space encloses the first $\tilde n$ poles. $U$ can be any function that does not obstruct the  contour with singularities. 
The identity \eqref{eq:sumtoint} is a useful link between the KK and closed form representations of the propagator.

\section{Interactions: Dimensional Analysis } 
\label{se:NDA}

A key ingredient of our study is the magnitude of the couplings of the bulk scalar from an effective field theory (EFT) perspective. In the presence of interactions, a five-dimensional theory is understood to be an EFT with some ultraviolet cutoff $\Lambda$ beyond which the EFT becomes strongly coupled.
This cutoff is tied to the strength of interactions through dimensional analysis in the strong coupling limit through so-called na\"ive dimensional analysis (NDA)~\cite{Manohar:1983md, Georgi:1986kr, Georgi:1992dw, Chacko:1999hg, Panico:2015jxa}; see e.g.~\cite{Ponton:2012bi} for a pedagogical introduction of NDA to 5D theories.
The crux of this analysis is to compare amplitudes of different loop order or involving higher dimensional operators. Let us define the loop factors 
\begin{align}
	\ell_5 &=24\pi^3
	&\text{and}&&
	\ell_4&=16\pi^2 \ .
\end{align}

\subsection{Gravitational Interactions}

The interactions of the graviton in AdS is controlled by the dimensionless coupling
\begin{align}
	\kappa &= \frac{k}{M_{\rm Pl}} \,. 
	\label{eq:kappa}
\end{align}
The reduced 4D and 5D Planck masses are related by $M_5^3=M^2_{\rm Pl}k$. 
By NDA, the cutoff in the gravity sector
\begin{align}
	\Lambda_\textnormal{grav}^3 &= \ell_5 M_5^3= \ell_5\kappa M_\textnormal{Pl}^3 \ .
	\label{eq:lambda_grav}
\end{align}
In order to keep  higher order gravity terms under control, $\kappa$ should be at most $\mathcal O(1)$~\cite{Chacko:1999hg,Agashe:2007zd}.

The gravity cutoff $\Lambda_\textnormal{grav}$ is sometimes taken as a universal scale setting the strength of all interactions in the effective Lagrangian. However, in the EFT the typical strength of interactions in various sectors  can in principle be different with different strong coupling scales. Strongly-interacting matter cannot influence the strength of gravity, which is protected by diffeomorphism invariance and set by the background geometry. 
In particular, matter interactions are at least as strong as gravity. The strong coupling scale of pure matter interactions can thus be lower than  $\Lambda_\textnormal{grav}$. 
Notice that gravity can even be removed, $M_\textnormal{Pl}\rightarrow \infty$, while the matter cutoff remains unchanged.\,\footnote{In a UV completion, the $\Lambda$, $\Lambda_{\rm grav}$ scales would likely be correlated and a fine-tuning might be needed to separate these scales.   }

\subsection{Matter Interactions \label{se:matter}}

We assume that a universal cutoff $\Lambda$ sets the strength of interactions in the matter sector of our theory.
To make this connection manifest in $D$-dimensions, one writes the fundamental action in terms of dimensionless fields $\hat \Phi$ with $\ell_D$ factored out~\cite{Chacko:1999hg, Ponton:2012bi}:
\begin{align}
	S_D &= 
	\frac{N_s\Lambda^D}{\ell_D} 
	\int d^D X \hat {\cal L}\left[\hat \Phi, \partial/\Lambda\right]
	\,.
	\label{eq:SD_NDA}
\end{align}
$N_s$ counts the number of species in the Lagrangian; for the present study we set $N_s =1$.
NDA states that an $\mathcal O(1)$ coupling in $\hat {\cal L}$ corresponds to a strong interaction strength. 
The dimensionful Lagrangian is recovered by canonically normalizing the fields. 
For the case of a cubic interaction, the NDA coupling dictated by \eqref{eq:SD_NDA} is $\lambda\sim (\ell_5\Lambda)^{1/2}$. 

The gravitational cutoff $\Lambda_\textnormal{grav}$ is related to the AdS curvature $k$ through \eqref{eq:kappa} and \eqref{eq:lambda_grav}. One may determine a similar relation between the matter cutoff $\Lambda$ and $k$ by considering the effective 4D interactions between specific KK modes. 
When expanding the 5D field in terms of canonically normalized 4D modes, $\Phi= kz \sum_n \tilde f_n(z) \phi_n(x)$, one finds that $\tilde f_n(1/\mu)$ is of order $ \sqrt{k}$.\,\footnote{
The KK mode normalization is $\int dz (kz)^{-1}\tilde f_n(z) \tilde f_m(z)= \delta_{mn}$. One has $\tilde f_m(z)= (kz)^{-1} f_m(z) $, where the $f_m$ are introduced in Sec.~\ref{se:AdS}. }
Because KK modes are localized towards the IR brane, this implies that the order of magnitude of an effective 4D coupling between KK modes is obtained from the 5D coupling by multiplying by powers of $\sqrt{k}$ and the warp factor $w=\mu/k$. 
For a given KK mode, the 4D NDA action is
\begin{align}
S_\textnormal{KK} = \frac{w^4 \Lambda^4 }{\ell_4} \int d^4 x \hat {\cal L}\left[\hat \phi, \partial/(w \Lambda )\right]\,
\label{eq:SD_KK}
\end{align}
following the same conventions of~\eqref{eq:SD_NDA}.  Notice that the cutoff only appears through the warped down cutoff scale $w \Lambda  = \tilde \Lambda$; we discuss this feature in Section~\ref{se:transition}.

Consider a general monomial interaction $\lambda_\textnormal{5D} \Phi^n/n!$ in the 5D action with $n>2$. 5D NDA, \eqref{eq:SD_NDA}, reveals that the strong coupling coefficient is 
\begin{align}
	\lambda_\textnormal{5D}={\ell_5}^{n/2-1} \Lambda^{5-3n/2}\,.
\end{align}
An interaction between $n$ KK modes with $\mathcal O(1)$ dimensionless couplings is then
\begin{align}
	\lambda_\textnormal{4D} \sim {\ell_5}^{n/2-1} \Lambda^{5-3n/2} k^{n/2-1} w^{4-n}\,. 
	\label{eq:gam4_matched}
\end{align}
On the other hand, the 4D NDA value for $\lambda_4$ is
\begin{align}
	\lambda_4={\ell_4}^{n/2-1} \Lambda^{4-n}w^{4-n}\,.  
	\label{eq:gam4_NDA}
\end{align}
For the effective theory of KK modes to be valid, one must require the effective $\lambda_4$ in \eqref{eq:gam4_matched} to be smaller than or equal to its strong coupling estimate, \eqref{eq:gam4_NDA}. 
This implies 
\begin{align}
	\Lambda > \frac{\ell_5}{\ell_4}k \ .
	\label{eq:Lambda_bound}
\end{align}
This universal relation arises because the $\sqrt{k}$ and the loop factors have the same powers in the NDA estimates, which are in turn fixed by field counting. 
When \eqref{eq:Lambda_bound} is not saturated, the effective 4D couplings of KK monomials are suppressed by powers of $(\ell_5 k/ \ell_4\Lambda)^{1/2}$ with respect to their strong coupling value.   This systematic suppression factor is reminiscent of the large $N$ suppression in the dual CFT, see Section \ref{se:ndaadscft}.

\subsection{Value of the Cubic Coupling \label{se:cubic}}

In this work we consider a scalar field,  whose natural mass scale would be $\mathcal O(\Lambda)$, as reflected by NDA. 
While the NDA value of the cubic coupling is $\lambda \sim (\ell_5 \Lambda)^{1/2}$, for this manuscript we set it to a smaller value 
\begin{align}\lambda \sim m_\Phi \frac{\ell_5^{1/2} }{\Lambda^{1/2}}\,.
 \label{eq:lambda_NDA} 
\end{align}
This value is consistent with a bulk mass parametrically lower than $\Lambda$:
 the self-energy bubble diagram from $\lambda$ gives a $\mathcal O(m_\Phi^2)$ contribution, in accordance with NDA.
The $\lambda$ coupling tends to zero in the free limit $\Lambda\rightarrow \infty$ (i.e.~$N\rightarrow \infty$) as it should.

\section{The Kaluza--Klein and Continuum Regimes of AdS}
\label{se:regimes}

We study the behavior of the effective theory using the results of the free theory in Section~\ref{se:AdS} and the interaction strengths in Section~\ref{se:NDA}.
Quantum corrections from the bulk interactions `dress' the bulk propagator and cause it to have qualitatively different behavior depending on the four-momentum, $p$. We show how these corrections separate the Kaluza--Klein and continuum regimes of a bulk scalar. 

\subsection{The Transition Scale \label{se:transition}}

The homogeneity of AdS implies a homogenous 5D cutoff on proper distances smaller than $\Delta X \sim 1/\Lambda$. In the conformal coordinate system the cutoff is $z$-dependent with respect to the Minkowski distance, since $\sqrt{\eta_{\mu\nu}\Delta x^\mu\Delta x^\nu}\sim kz/\Lambda $.
In position--momentum space the condition amounts to $p\sim \Lambda/(kz) $. 
This implies that the 5D cutoff for an observer at position $z$ in the bulk is warped down to $\Lambda/(kz)$. 

One can see this from an EFT perspective: the effects of higher-dimensional operators in the action are enhanced by powers of $z$. For example, consider dressing the propagator with a higher derivative bilinear, $\square(\partial_\mu \Phi)^2/\Lambda^2$ with an $\mathcal O(1)$ coefficient as dictated by NDA (see Eq.\,\eqref{eq:SD_NDA}). This term dominates for
\begin{align}
	p z\gtrsim \Lambda /k\,. \label{eq:cutoff}
\end{align}
For a fixed $p$, this implies that the EFT breaks down in the IR region of AdS, $z \gtrsim (\Lambda/k)/p$; see {e.g.}~\cite{ArkaniHamed:2000ds,Goldberger:2002cz, Fichet:2019hkg}. 
The cutoff is warped below the scale $p$ for values of $z$ beyond this region. Therefore propagation into this region of position--momentum space falls outside the EFT's domain of validity.

It follows that the theory also contains a scale

\begin{align}
\tilde{\Lambda} = \Lambda \frac{\mu}{k}\,,
\end{align}
the warped down cutoff at the IR brane. At energies $p>\tilde \Lambda$, the correlation functions cannot know about the IR brane since it is in the region of position--momentum space hidden by the EFT validity condition~\eqref{eq:cutoff}.
In short, for $p>\tilde \Lambda$ the IR brane is ``outside of the EFT,'' see Section~\,\ref{se:picture}.

This is a hint that the behavior of the theory undergoes a qualitative change at $\tilde \Lambda$. The IR brane imposes a boundary condition that leads to discrete KK modes. Thus for $p<\tilde \Lambda$, one can expect that the theory features KK modes. On the other hand, for $p>\tilde \Lambda$ the IR brane is outside the EFT, hence no KK modes should exist. Instead, an observer should see a continuum of states.

\subsection{Dressed Propagator \label{se:opacity}}

The free propagator in \eqref{eq:propa} encodes narrow KK modes. It amounts to $\Lambda \rightarrow \infty$ or $N \rightarrow \infty$. 
The continuum behavior becomes apparent when one dresses the free propagator with quantum corrections.\,\footnote{
    The exact calculation of diagrams in AdS has recently been an intense topic of research, see {e.g.}~\cite{ Aharony:2016dwx,Yuan:2017vgp,Giombi:2017hpr,Carmi:2018qzm,Carmi:2019ocp,Meltzer:2019nbs} for loop-level diagrams and \cite{Bzowski:2013sza,Isono:2018rrb,Albayrak:2019asr, Albayrak:2020isk} for developments in position--momentum space. 
    Throughout this paper we instead use approximate propagators.
    }
These quantum corrections resolve the poles in the free propagator with timelike momenta as they do in 4D Minkowski space.
Including these effects corresponds to evaluating the leading $1/N^2$ effect on the propagator of the strongly coupled dual theory; in our case this is $1/N^2\sim\lambda^2/k$.

We focus on bulk self-energy corrections from a cubic self-interaction. Brane-localized self-energies only modify the boundary conditions and are thus unimportant for our purposes. 
In contrast to the free propagator, the Green's function equation for the dressed propagator satisfies
\begin{align}
	D_X\Delta(X,X') 
	- \frac{1}{\sqrt{g}}\int dY \,\Pi(X,Y) \Delta\,(Y,X')
	=
	-\frac{i}{\sqrt{g}}\delta^{(5)}(X-X')
	\,, 
	\label{eq:DDelta_dressed}
\end{align}
where $i\Pi(X,Y)$ are 1PI insertions that dress the propagator. In our case, the leading $i\Pi$ insertion is induced by the scalar bubble induced by the $\lambda \Phi^3$ interaction. We are interested only in the imaginary part of the self-energy, which is finite.

A calculation of $i\Pi(X,Y)$ is performed analytically in \cite{Fichet:2019hkg} with self-consistent approximations in the limit of strong coupling and moderate bulk masses $\alpha=\mathcal O(1)$.  
One of the tricks for the analytical estimate is to expand the non-local self-energy as a series of local insertions, which  amounts to a $\partial_z$ expansion.
Using this method, we estimate of the contribution from the  $|p|> 1/z_<$ regime. The imaginary part of the 1-loop bubble induces a shift of $p$, 
\begin{align}
	\Delta_p^{\text{dressed}}(z,z')
	&\sim\Delta^{\text{free}}_{p(1+ic)}(z,z')
	&
	c&\sim a \frac{\lambda^2}{ \ell_5  k}
	\,,
	\label{eq:dressing}
\end{align}
where $c$ is loop-induced and estimated to have $a\sim \mathcal O(1/10)$ with a large uncertainty.
\footnote{ This estimate  is confirmed in the upcoming detailed analysis of \cite{Costantino:2020vdu}. 
}
Using the NDA value of $\lambda$ in \eqref{eq:lambda_NDA} and taking $m_\Phi=\mathcal O(k)$, one finds $c \sim a k /\Lambda \sim a/(\pi N^2) $. The  $|p|> 1/z_<$ regime provides a larger contribution to $a$ than the result previously presented from the $|p|< 1/z_>$ regime \cite{Fichet:2019hkg}. This extends the validity of our calculations to weaker coupling, hence allowing large $N$. A self-consistent numerical solution to the integro-differential equation of motion, \eqref{eq:DDelta_dressed}, may be required to obtain the general dressed propagator. We leave this for future work.

\subsection{The Two Regimes \label{se:regimes_sub}}

The self-energy dressing of the propagator presents distinct Kaluza--Klein and continuum regimes.
The poles of the free propagator are set by zeros of its denominator. 
For momenta much larger than the IR brane scale, $p\gg \mu$, the asymptotic form of the Bessel functions lead to a propagator that is approximately proportional to   
\begin{align}
	\Delta_p(z,z')
	&\propto
	\frac{1}{\sin\left(\frac{p}{\mu}-\frac{\pi}{4}\left(1+2\alpha\right)\right)}\ .
\end{align}
The effect of the dressing, \eqref{eq:dressing}, softens the poles and causes them to merge at a scale 
\begin{align}
	p
	&\sim 
	\frac{\mu}{c} \sim \frac{\tilde \Lambda}{a} 
	\label{eq:pole_cond}  \, .
\end{align} 
Above this scale the propagator describes a continuum rather than distinct Kaluza--Klein modes.
Thus we observe that the dressing of the propagator reaffirms the existence of distinct KK and continuum regimes separated by a transition scale controlled by $\tilde \Lambda=(\mu/\Lambda){k}$.
Let us comment further on both sides of the transition.

\subsection{Kaluza--Klein Regime: \texorpdfstring{$p<\tilde \Lambda$}{p < Lambda-tilde}}
\label{sec:KK:regime}

For momenta less than the transition scale $\tilde\Lambda$, {UV correlation functions are} sensitive to the physics of the IR brane.
The IR brane provides a boundary condition for the bulk equation of motion and hence imposes a discrete spectrum of KK modes. These modes may be narrow. However, as the KK mass approaches the transition scale, the KK modes must merge to form a continuum. To see this, one may use the full form of the dressed KK propagator 
from \eqref{eq:DDelta_dressed}. This propagator is given by
\begin{align}
	\Delta_q(z,z')
	=
	i\,\mathbf{f}(z) 
	\cdot 
	\left[ \mathbf{D}^{-1} + i\,\textrm{Im}\,\mathbf{\Pi} \right]^{-1}
	\cdot  \mathbf{f}(z') \,
\end{align}
where 
\begin{align}
	i \mathbf{\Pi} 
	&\equiv
	\int du \int dv \, i\Pi(u,v) \mathbf{f}(u) \otimes \mathbf{f}(v)
	 \,.
\end{align}
The imaginary part of $\mathbf{\Pi}$ gives rise to a ``width matrix'' for the KK resonances. Critically, $\mathrm{Im}\,\mathbf{\Pi}$ is \emph{not} diagonal: the KK modes mix due to this non-diagonal, imaginary contribution to the mass matrix. 
The KK modes may merge into a continuum either because they become broad, or because of the mixing induced by $\mathrm{Im}\,\mathbf{\Pi}$.
This property of the AdS propagator is suggestive of how heavy glueballs in the strongly-coupled dual tend to merge near the $\tilde \Lambda$ cutoff, see Section~\ref{se:CFT_glueballs}.

At low enough four-momentum $p$, the narrow-width approximation applies to the KK modes. The KK modes can then be treated as asymptotic 4D states. The optical theorem applies to these light KK modes. In contrast, when approaching the transition scale, the KK modes cannot be seen as asymptotic states due to large widths and KK-mode mixing. This is consistent with the properties of non-truncated AdS.

\subsection{Continuum Regime: \texorpdfstring{$p > \tilde \Lambda$}{p>Lambda-tilde}}

When $p$ is above the transition scale, $\tilde \Lambda$, the oscillating pieces of the propagator are smoothed. Within this regime, the endpoints of the propagator define additional scales for which the propagator realizes different behavior.

\paragraph{Continuum regime, low momentum.}
In the continuum regime with low momentum, $|p|>\tilde \Lambda$ and $|p| < z_>^{-1}$, and away from the poles, the propagator is
\begin{align} 
	\label{eq:propaII}
	\Delta_p(z,z') 
	&\approx
	\Delta_\UV + \Delta_\textnormal{heavy} + \Delta_\textnormal{light}
	\ ,
\end{align}
where the pieces are
\begin{align} 
 	\Delta_\UV
 	&=
 	\phantom{+} i \frac{(b_{UV}+2\alpha)(kz)^{2-\alpha}(kz')^{2-\alpha}}{\alpha \left(p^2/(\alpha-1)k+ 2b_{UV}k\right)}
	&
	\Delta_\textnormal{heavy} 
	&= 
	-i\frac{(kz)^2(kz')^2}{2\alpha k} \left(\frac{z_<}{z_>}\right)^{\alpha}
	\\
	\Delta_\textnormal{light}
	&=
	-i \frac{\Gamma(-\alpha)(kz)^2(kz')^2}{\Gamma(\alpha+1)2b_{UV}^2k}
	g(z_<)
	g(z_>)
	\left(
		\frac{-p^2}{4k^2}
	\right)^\alpha 
	&
	g(z) 
	&= 
	\frac{b_\UV+2\alpha}{(zk)^\alpha} -b_{\UV}(zk)^\alpha
	\, .
\end{align}
Notice that the dependence on the $\mu$ parameter has dropped this expression. This is a manifestation of the propagator's agnosticism of the IR brane in this regime.
Conversely, this implies that when varying $p$ from UV scales to IR scales, the IR brane is effectively emergent when $p$ drops below $\tilde \Lambda$.

The content of each term in \eqref{eq:propaII} is also instructive. The first term, $\Delta_\UV$ represents a 4D  mode localized near the UV brane.\,\footnote{In our convention, the 4D mode squared mass is positive for negative $b_{\UV}$.}
This 4D mode is assumed to be very heavy, $b_{\UV}=\mathcal O(1)$, such that it does not play a role in the processes of in this manuscript. The second term, $\Delta_\textnormal{heavy}$ is analytic and encodes the collective effect of heavy KK modes. The third term, $\Delta_\textnormal{light}$ is nonanalytic and encodes the collective effect of light modes.

\paragraph{Continuum regime, high momentum.}
In the continuum regime with high momentum, $|p| >\tilde \lambda$ and $z_>^{-1}<|p|<z_<^{-1}$,  the numerator of the propagator oscillates:
\begin{align}
	\Delta_p(z,z') 
	&\propto 
	\frac{\cos\left({p}{\mu^{-1}}-pz_>\right)}{\cos\left({p}{\mu^{-1}}+\varphi_-\right)}
	\times
	\begin{cases}  
		\displaystyle
        1
		&
		\textnormal{for}~{z_>^{-1}}<p<{z_<^{-1}}
		\\[1em]
		\displaystyle
			\cos\left(pz_<-\varphi_+\right)
		&\textnormal{for}~p>{z_<^{-1}} 
		\ ,
 \end{cases}
 \end{align}
where we have written phase shifts as $\varphi_\pm={\pi}\left(1\pm2\alpha\right)/4$. Upon dressing, the non-oscillatory part of the propagator in this region scales as
\begin{align}\Delta_p(z,z') \sim \begin{cases}  
e^{- |p| z_>} \quad &\textnormal{for}~p_\mu~\;\textnormal{spacelike}
\\
 e^{-c p z_>} \quad &\textnormal{for}~p_\mu~\;\textnormal{timelike}
 \end{cases} 
 \label{eq:Kexp} \ .
\end{align} 
This is an important feature: the IR region of AdS is opaque to propagation for both spacelike and timelike momenta. 
The regions of opacity are somewhat different---the suppression for spacelike momentum occurs at $z \sim 1/|p|$, while the suppression for timelike momentum occurs at $z \sim 1/{cp}$.
Substituting in $c$, we see that the suppression in the timelike regime occurs for 
\begin{align}
	pz_> \gtrsim \frac{\Lambda}{a k}\,. \label{eq:op}
\end{align}
This behavior is similar to the region of EFT breaking in \eqref{eq:cutoff}. Therefore the opacity of the space effectively censors the region where the EFT breaks down. This behavior was qualitatively predicted in Ref.~\cite{ArkaniHamed:2000ds}. 
For the specific case with an endpoint on the IR brane, $z_>=1/\mu$, the opacity threshold \eqref{eq:op} is the same as the scale at which the Kaluza--Klein poles disappear, \eqref{eq:pole_cond}. The two effects are, of course, closely related: the poles vanish precisely when the IR brane becomes opaque to the propagator. 

\paragraph{} 
In the continuum regime, KK modes are not appropriate variables to describe the theory because the $f_n$ profiles fall into a spacetime region where the EFT breaks down, \eqref{eq:cutoff}.  Instead, the meaningful variables are those localized on the UV brane. These remain in the theory up to the ultimate cutoff $p\sim \Lambda$. This was already observed in \cite{Goldberger:2002cz} from EFT considerations, and is completely consistent with the holographic formalism needed for AdS/CFT.

\section{Cascade Decays in the Continuum Regime}
\label{se:Decays}

The same bulk interactions that induce opacity in the IR region necessarily induce cascade decays in the bulk.
These cascade decays, in turn, may appear to be a possible loophole to the arguments in the previous section.
In particular, it is  possible that a continuum with $p\gg \tilde \Lambda$  undergoes cascade decays down the KK regime, ending in light narrow KK states and/or in IR-localized states.
In such a process, it may seem that for \emph{any} initial momentum the cascade decay  `knows' that an IR brane exists. 
This appears to circumvent the  picture obtained in Section~\ref{se:regimes}, where the theory at $p\gg \tilde \Lambda$ does not know at all about the IR brane.
We evaluate explicitly this process in this section and discuss implications in Section~\ref{se:picture}.

The properties of cascade decays initiated in the KK regime are fairly well-understood and are summarized in Section~\ref{se:shape}. We instead focus on the cascade decays starting in the continuum regime. This regime is always present unless interactions are removed ($\Lambda\rightarrow \infty$). Furthermore, in the strong coupling limit $\Lambda\sim k$, there is essentially no KK regime and all propagation is in the continuum regime.  We seek to determine the overall shape and the total probability for a cascade decay event to occur in the continuum regime. 

The bulk of AdS does not permit asymptotic states or a conventional $S$-matrix  (see {e.g.} \cite{Giddings:1999qu,Balasubramanian:1999ri}). However the 4D modes localized on the branes, which have a 4D Minkowski metric, can provide usual asymptotic states. 
We thus consider decays that are initiated on the UV brane. The decay may end back on the UV brane or reach asymptotic states on the IR brane. It can also end in narrow KK modes which are effectively asymptotic states in the limit of the 4D narrow width approximation.

\subsection{The Decay Process}

The explicit evaluation of a generic decay diagram with an arbitrary number of legs is, in principle, challenging because there are many phase space and position integrals to perform over a non-trivial integrand. 
However, it turns out that a recursive approach can be adopted based on simplifying approximations. We build on this approach to estimate the total rate for a generic decay.

For intermediate steps in this calculation, it is convenient to formally write the final states as KK modes, even if the corresponding momenta are in the continuum regime. Sums over KK modes may then be re-expressed in terms of the closed form propagator at the end of the calculation. 

Measurable event rates, such as cross sections and decay widths, depend on the integral of the squared amplitude over phase space.
To emphasize that our approach does not depend on how the continuum is created, we work at the level of this integrated square amplitude, denoted as $P_M$. 
\begin{figure}[t]
	\includegraphics[width=1.00\linewidth]{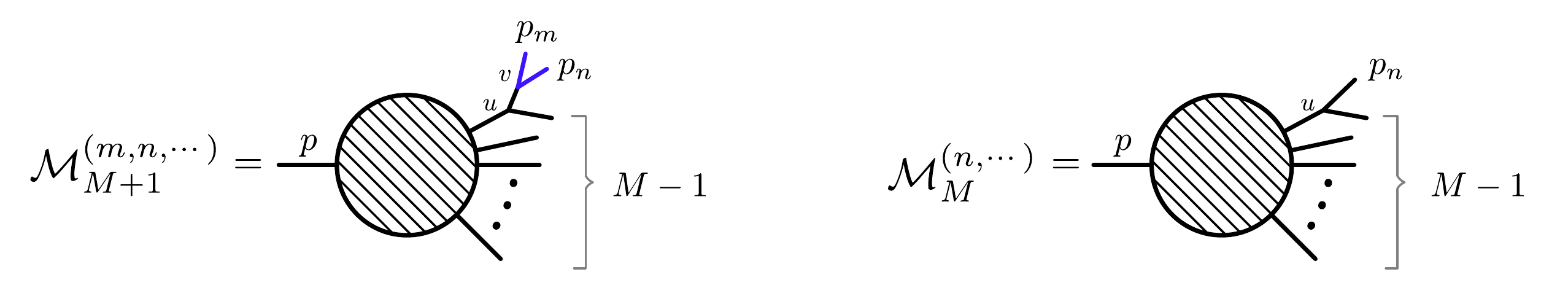}
		\caption{The cascade decay amplitudes. $u$ and $v$ are coordinates in the $z$ direction. In our recursive approach, we relate the integrated square amplitude of the left diagram to that of the right diagram. \label{fig:amps} }
\end{figure}
For the diagram in Fig.~\ref{fig:amps} with $M+1$ final states,
\begin{align}
	P_{M+1} \equiv 
	\int\sum_{\textnormal{FS}(M+1)}|\mathcal{M}_{M+1}|^2(2\pi)^4d\Phi_{M+1}.
\end{align}
The sum over ${\textnormal{FS}(M+1)}$ is shorthand for a sum over all possible combinations of $(M+1)$ KK modes that are kinematically allowed final states. 
$d\Phi_{M+1}$ is the volume element of the $(M+1)-$body Lorentz-invariant phase space~\cite{PhysRevD.98.030001}.
We label specific specific final state KK numbers and four-momenta as $m$, $p_m$ and $n$, $p_n$.
The amplitude for a given set of final state KK modes is expressed as 
\begin{align}
	\mathcal{M}_{M+1}^{(m,n,\cdots)} 
	&= 
	\int du ~ \mathcal{I}_{M}^{(\cdots)}(u)
	\int_{1/k}^{1/\mu} dv \frac{\lambda\Delta_{q}(u,v)}{(kv)^5}f_{m}(v)f_{n}(v)
	\ .
	\label{eq:amp}
\end{align}
$\mathcal{I}_{M}^{(\ldots)}(u)$ is the amplitude that has been amputated just before the propagator that produces the $m$ and $n$ modes, see Fig.~\ref{fig:amps}.

The $\mathcal{M}_M$ amplitude, shown on the right-hand side of Fig.~\ref{fig:amps}, is
\begin{align}
	\mathcal{M}_{M}^{(n,\cdots)} = \int du ~\mathcal{I}_{M}^{(\cdots)}(u)f_{n}(u).
\end{align}
The corresponding integrated square amplitude is 
\begin{align}
	P_{M}\equiv\int\sum_{FS(M)}|\mathcal{M}_{M}|^2(2\pi)^4d\Phi_{M}
	\ .
\end{align}
We now relate $P_{M+1}$ to $P_M$. 

\subsection{Recursion Relation}

\label{se:recursion}

Propagators with timelike momentum are suppressed beyond $z_>\sim 1/( c p)$, as seen in \eqref{eq:Kexp}. We assume for simplicity that $c\sim 1$. This implies that our evaluation assumes nearly strong coupling, {i.e.}~$\Lambda$ is not far from $k$. Following this, the position integrals effectively have no support beyond $z\sim 1/p$. Note that this is equivalent to only considering contributions from the  $\mu<|p|<z_>^{-1}$ region of position--momentum space, see \eqref{eq:propaII}. 

We have numerically evaluated contributions from the $|p|>z_>^{-1}$ regions and found that they tend to be smaller or of the same order as the results from this section for $c$ near unity. These contributions can be somewhat larger for smaller $c$, though a detailed analysis is beyond the scope of this manuscript.

We square the amplitude and write sums on KK modes as integrals over the propagator using \eqref{eq:sumtoint}. In the continuum regime, only the third term of the continuum propagator in \eqref{eq:propaII} contributes to the contour integral because it carries a branch cut. By deforming the contour to fit snugly around the branch cut, we determine that
\begin{align}
	\sum_{n=0}^{\tilde n} U(m_n^2) f_n(z) f_n(z')
	=
	-\frac{1}{2\pi}
	\oint_{\mathcal C \left[\tilde n\right]}
	dq^2 \, U(q^2) \Delta_q(z,z')
	=
	\frac{-1}{2\pi}
	\int_0^{m_{\tilde{n}}^2}
	dq^2 \, U(q^2) \mathrm{Disc}[\Delta_{q}(z,z')]
	\label{eq:sumtoint2} \, .
\end{align}
In terms of the propagator, $P_{M+1}$ then reads
\begin{alignat}{3}
P_{M+1} 
	&= 4\pi^2 \sum_{\textnormal{FS}(M-1)} 
	&& \int d\Phi_{M+1} \int du \int  du'\,
	\mathcal{I}_{M}(u) \, \mathcal{I}^*_{M}(u')
	\int_{{1}/{k}}^{{1}/{q}}dv\label{eq:slowdecay}
	\int_{{1}/{k}}^{{1}/{q}}dv'~
	\frac{\lambda^2\Delta_{q}(u,v)\Delta_{q}^*(u',v')}{(kv)^5(kv')^5} \times
	\\
	& && 
	\int d\pmd^2\, \mathrm{Disc}[\Delta_{\pmd}(v,v')]
	\int d\pnd^2\, \mathrm{Disc}[\Delta_{\pnd}(v,v')]
	\, .
	\nonumber
\end{alignat}
The integrals over the $\pmd^2$, $\pnd^2$ variables implement the sum over KK modes in \eqref{eq:sumtoint2}. 
The integrands in \eqref{eq:slowdecay} carry positive powers of $v$ and $v'$ so that the $dv\, dv'$ integrand is largest at the upper limit, $v,\,v'\sim 1/q$. Because $q$ is the momentum flowing through the parent this implies that the cascade decay progresses slowly towards the IR region. 

We break up the phase space using the standard recursion relation, see e.g.~\cite{PhysRevD.98.030001},
\begin{align}
	d\Phi_{M+1}=d\Phi_2(q;\pmd,\pnd)\, d\Phi_M\, (2\pi)^3 dq^2
	\, .
\end{align}
We get
\begin{alignat}{3}
P_{M+1} 
	&=(2\pi)^4 \sum_{\textnormal{FS}(M-1)} 
	&& \int d\Phi_{M}
	\int du \int  du'\,
	\mathcal{I}_{M}(u) \, \mathcal{I}^*_{M}(u')
	\int_{{1}/{k}}^{{1}/{q}}dv\label{eq:integrand}
	\int_{{1}/{k}}^{{1}/{q}}dv'~
	\frac{\lambda^2}{(kv)^5(kv')^5} \times
	\\
	& && 
	\int d\pmd^2\, \mathrm{Disc}[\Delta_{\pmd}(v,v')]
	\int d\pnd^2\, \mathrm{Disc}[\Delta_{\pnd}(v,v')]
	\int\frac{dq^2}{64\pi^4q^2}K(q,\pmd,\pnd)\Delta_{q}(u,v)\Delta_{q}^*(u',v')
	\, .
	\nonumber
\end{alignat}
Here $K(q,\pmd,\pnd)$ is the 2-body kinematic factor,
\begin{align}
	K(q,\pmd,\pnd)^2
	\equiv
	{\left[
		q^2-\left(\pmd+\pnd\right)^2
	\right]
	\left[
		q^2-\left(\pmd-\pnd\right)^2
	\right]} \ .
\end{align}
We approximate the integrals over $\pmd^2$ and $\pnd^2$ as
\begin{align}
	\int_0^q d\pmd \int_0^{q-\pmd} d\pnd
	~\pmd^{2\alpha+1}\pnd^{2\alpha+1}
	K(q,\pmd,\pnd)
	\approx
	\int_0^{{q}/{2}}
	d\pmd 
	\int_0^{{q}/{2}} d\pnd~ \pmd^{2\alpha+1}\pnd^{2\alpha+1} q^2\,.
\label{eq:kinapprox}
\end{align}
This approximation introduces a ${\cal O}(1)$ error that depends on $\alpha$.\footnote{The error monotonically increases from $\sim25\%$ for $\alpha$ near 0 to $\sim30\%$ for $\alpha$ near 1.} Note that the dominant contribution to the integral in \eqref{eq:kinapprox} comes from the region near the upper limit. This indicates that the continua tend to decay near kinematic threshold. Thus the cascades gives rise to soft spherical final states, in accordance with former results from both gravity and CFT sides.

Integrating over $\pmd^2$, $\pnd^2$, $v$, and $v'$, we have
\begin{align}
	P_{M+1}
	=&
	C_\alpha
	\sum_{FS(M-1)}(2\pi)^4
	\int d\Phi_M
	\int \frac{dq^2}{k}
	\left(\frac{q}{k}\right)^{2\alpha}
	\int du 
	\int du'
	\, \mathcal{I}_{M}(u)
	\, \mathcal{I}_{M}^*(u')
	(ku)^{2+\alpha}(ku')^{2+\alpha}\,, 
\end{align}
where the constant prefactor is
\begin{align}
	C_\alpha &= 
		\frac{
			8^{4(1-\alpha)}\lambda^2 
		}{
			\alpha^4\pi^4k
		}
	\left(
		\frac{\Gamma(1-\alpha)\sin(\pi\alpha)}{\Gamma(1+\alpha)}
	\right)^2
	\frac{
		|(2+3 \alpha)4^\alpha -(\alpha +2)
		\frac{\Gamma(1-\alpha)}{\Gamma(1+\alpha)} 
		e^{i\alpha\pi}|^2
	}{
		(2+3\alpha)^2 (2+ \alpha )^2 (1+\alpha)^2
	} \ .
\end{align}
One may replace the $dq^2$ in favor of a sum over the continuum of KK final states by applying \eqref{eq:sumtoint2}. This yields a recursion relation
\begin{align}
	P_{M+1}
	= r \int\sum_{\textnormal{FS}(M)}
	\left|\int  du ~ \mathcal{I}_{M}(u)f_{n}(u)\right|^2
	(2\pi)^4d\Phi_n 
	= r \, P_M
	\ . \label{eq:rec}
\end{align}
The fact that one obtains a simple relation is a consequence of the integrand having a specific momentum dependence and is nontrivial. This relation is clearly useful since it can be used to give an estimate of a total rate with arbitrary number of legs. 

The recursion coefficient $r$ is given by
\begin{align}
	r \equiv
	\frac{\lambda^2}{k}
	\frac{1}{1024^{1+\alpha}}
	\frac{1}{2\pi^3\alpha^3}
	\left(
		\frac{
			|(2+3 \alpha)4^\alpha -(2+\alpha )
				\frac{\Gamma(1-\alpha)}{\Gamma(1+\alpha)} e^{i\alpha\pi}|^2
		}{ 
			(2+3\alpha)^2 (2+ \alpha )^2 (1+\alpha)^2
		}\right)
		\frac{\Gamma(1-\alpha)\sin(\pi\alpha)}{\Gamma(1+\alpha)}
		\ .
		\label{eq:recursion}
\end{align}
Even for the strongly coupled case, $\lambda^2 \sim \ell_5 k$, this coefficient is much smaller than one. 

\section{Soft Bombs and the Emergence of the IR Brane}
\label{se:picture}

The recursion relation \eqref{eq:recursion} allows us to study the qualitative features of a complete cascade decay event. An event initiated on the UV brane with timelike momentum $P>\tilde \Lambda$ starts in the continuum regime and decays as a cascade of continua. This decay eventually reaches the KK regime.

\subsection{Shape\label{se:shape}}

The differential event rate---the integrand in the expression for $P_M$---determines the most likely configurations in phase space. The phase space approximation \eqref{eq:kinapprox} shows that decays tend to occur near threshold with final momenta evenly split between the offspring. 
The event thus tends to be soft and spherical. This confirms the \emph{soft bomb} picture obtained in the KK regime \cite{Csaki:2008dt}, in string calculations (see {e.g.}~\cite{Hofman:2008ar}) and in the gauge theory dual~\cite{Hatta:2007cs,Hatta:2008tn}.

The integrand in \eqref{eq:integrand} shows that vertices tend to occur at $z\sim 1/p$ where $p$ is the momentum of the parent continuum. 
There is a sense of progression in the fifth dimension: the cascade decay proceeds from the UV to the IR with each offspring moving further into the IR than its parent. 

\begin{figure}[t]
 	\centering
 	\includegraphics[width=\linewidth]{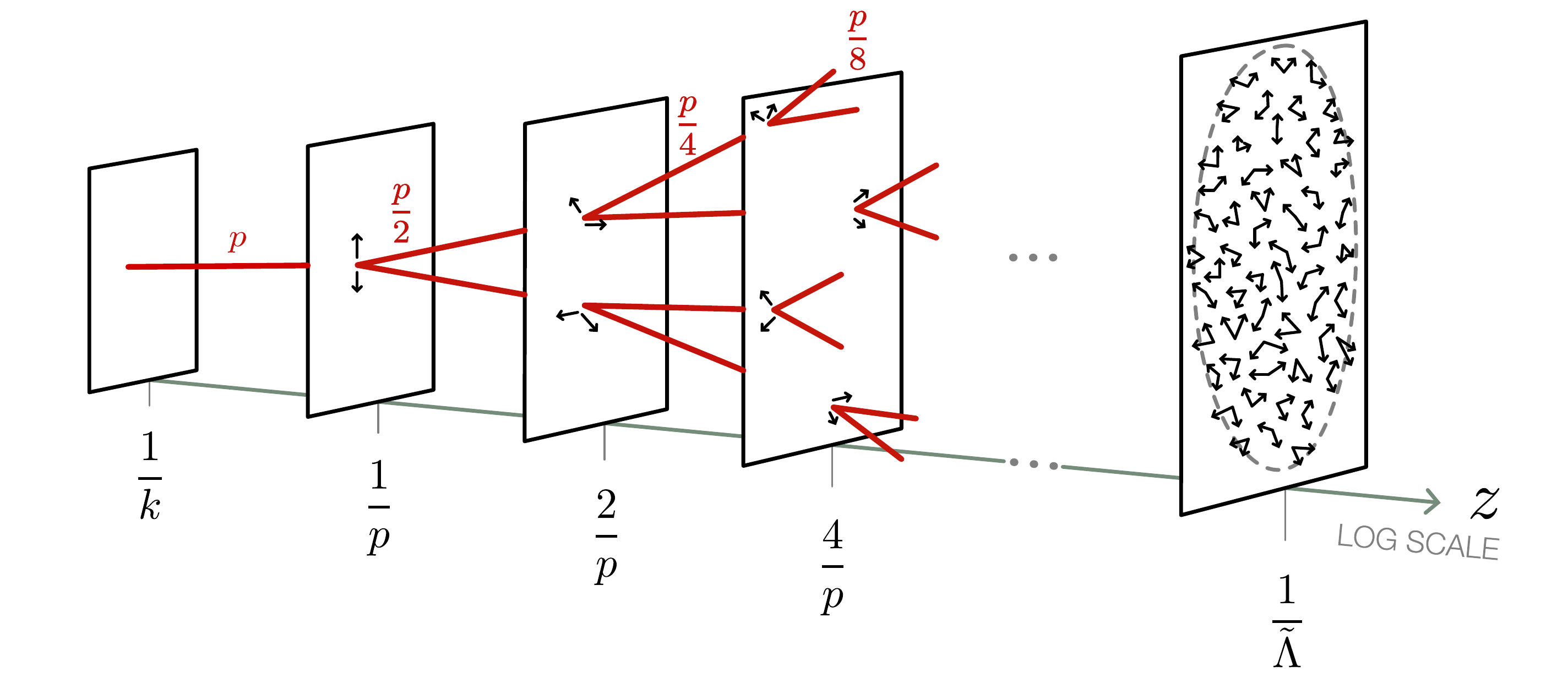}
 	\caption{ 
 	A typical field-theoretical soft bomb event in AdS$_5$ in the continuum regime $p>\tilde \Lambda$. The rate for such an event to occur is exponentially suppressed.
 	}	\label{fig:softbomb}
\end{figure}

Let $p_f$ be the average momentum of states after some number of branchings. The soft bomb then leaves the continuum regime and enters the KK regime at $p_f \sim \tilde \Lambda$. This is roughly the scale at which the KK modes become narrow. These features are summarized in Fig.~\ref{fig:softbomb}. 

\subsection{Total Rate} 

\label{se:rate}
 
The soft bomb enters the regime of narrow KK modes when the offspring  have average momenta $p_f \sim \tilde \Lambda$. At this scale, the narrow width approximation is valid and the recursion \eqref{eq:rec} halts because subsequent decays factorize. 
This highlights a key feature of the continuum regime in contrast to the KK regime: the phase space suppression factors are not compensated by narrow poles due to the breakdown of the narrow width approximation. This is why cascade events starting in  the continuum regime are suppressed. 

One can estimate the total rate of cascade decays 
using the recursion \eqref{eq:rec}.
A continuum cascade initiated with momentum $P$ stops at momentum $p_f\sim \tilde \Lambda$. Assuming an equal split of momenta among a total of $M$ offspring gives
\begin{align}
	M \sim P/\tilde \Lambda \ . \label{eq:mult}
\end{align}
The recursion relation \eqref{eq:rec} shows that the rate is suppressed by $r^{M-1}$.
\begin{align}
	P_M \sim r^{P/\tilde \Lambda} \,.
	\label{eq:suppr}
\end{align}
Since $r\ll 1$, the soft bomb is exponentially suppressed as a function of $P$ for initial timelike momenta in the continuum regime $P>\tilde \Lambda$.

 \subsection{Emergence of the IR Brane}

The suppression of the soft bomb rate in the continuum regime completes our picture of quantum field theory in AdS for timelike momenta. We can now make a statement about the `disappearance' of the IR brane in QFT first hinted in Section\,\ref{se:transition}.

Consider, for example, a UV-localized field $\varphi$ that couples to the bulk scalar, $\Phi$. The collision of two $\varphi$ states can induce a cascade decay $\varphi\varphi \rightarrow \Phi \rightarrow \Phi \Phi \rightarrow \cdots$
When the center-of-mass four-momentum is in the KK regime,  $P<\tilde \Lambda$,  the event rate is determined by the $\varphi\varphi \rightarrow \Phi^{(n)}$ amplitude to create an on-shell KK mode $\Phi^{(n)}$ with mass  $m_n \sim P$.
In contrast, in the continuum regime, $P>\tilde \Lambda$,  the cascade is initiated with 5D continua that have no poles and thus no notion of being on-shell. Narrow KK modes only appear after the cascade has produced enough offspring for the typical momentum to drop below $\tilde \Lambda$.
The amplitude  to calculate includes the entire cascade up to,  and including, the first narrow KK modes. 
The rate for a cascade in the continuum regime is  suppressed with respect to that in the KK regime by the tiny factor $r^{P/\tilde \Lambda}$ in \eqref{eq:suppr}.

This suppression implies that continua produced tend not to cascade down to many narrow KK states which can interact with an IR brane, but instead tend to go promptly into UV-brane states with no cascade. Thus in the continuum regime, the theory truly does not know about the IR brane.
The observables---including  decays---in this regime of the theory can be equivalently obtained in AdS background with \textit{no} IR brane.

Formally this statement can be spelled out using the partition function of the theory 
\begin{align}
e^{i E[J]}=\left<e^{ i    \int_{z_{\rm UV}}^{z_{\rm IR}} dz d^4p\, \Phi J} \right>=  \int{\cal D}[{\rm fields}] \exp{ i \left(   \int_{z_{\rm UV}}^{z_{\rm IR}} dz d^4p \left( {\cal L}_{\rm bulk} + \Phi J \right)+ S_{\rm UV}+ S_{\rm IR}\right)} \label{eq:Zfull}  \,,
\end{align}
where $E[J]$ is  the generating functional of the connected correlators. 
Our claim is that in the $p\gg \tilde \Lambda$ regime, the correlators are equivalently described by 
\begin{align}
e^{i E[J]\big|_{p\gg \tilde \Lambda}}\approx  \int{\cal D}[{\rm fields}] \exp{ i \left(   \int_{z_{\rm UV}}^{\infty} dz d^4p \left( {\cal L}_{\rm bulk} + \Phi J \right)+ S_{\rm UV}\right)}  \equiv e^{i E[J]\big|_{z_{\rm IR}\rightarrow \infty}}\,.
\label{eq:EFTUV}
\end{align}
On the right-hand side,  $ E[J]\big|_{z_{\rm IR}\rightarrow \infty}$ amounts to the theory with the IR brane removed. In other words, the IR brane---and the fields and operators localized on it---effectively vanishes for $p\gg \tilde \Lambda$. 
 Conversely, the IR brane affects correlators for lower $p$ and is thus effectively emergent.\,\footnote{
For the purpose of taking functional derivatives,  the source $J$ can formally be any distribution. 
If instead $J$ is given a physical meaning, it 
is typically localized towards the UV brane to avoid any backreaction of the metric towards the IR. 
In the context of holography, $J$ is exactly localized on $z=z_{\rm UV}$, giving rise to UV-localized variables as done in Section~\ref{se:ndaadscft}. }

Finally we notice that the continuum regime is exactly described by an appropriate CFT model as dictated by the AdS/CFT correspondence. Apart from the UV brane which amounts to a UV cutoff in the CFT, the theory is exactly AdS in the continuum regime.

\subsection{Optical Theorem}

In Section~\ref{sec:KK:regime} we observed that in the Kaluza--Klein regime, KK modes are valid asymptotic states that obey the optical theorem. 
In the continuum regime, even though the rate of cascade decays is exponentially suppressed, the imaginary part of the bulk self-energy ${\rm Im}\,\Pi$ is not. This does not contradict the optical theorem, though it may appear to do so when using the intuition from KK modes.
This is because unlike the KK regime, the continuum regime has no narrow state on which one may perform a unitarity cut. Thus the loop-level contribution to the self-energy is \emph{not} related to a decay---the optical theorem does not apply.

One may insist on identifying propagators of light KK modes with narrow widths upon which one may perform a unitarity cut. Because of the ``near-threshold'' property of KK vertices in Section~\ref{se:shape}, these light KK modes only appear at high loop order. A unitarity cut on this high-loop order diagram ultimately reproduces the typical soft bomb diagram in Figure~\ref{fig:softbomb} that ends in states with $m_{\rm KK}\sim \tilde \Lambda$. Such diagrams only amount to a tiny portion of  ${\rm Im}\,\Pi$.

\subsection{Asymptotically AdS Backgrounds \label{se:IR_gen} }

Our study focuses on a slice of pure AdS with no departure from AdS in the IR region. The qualitative features of our results can apply to models  whose backgrounds are deformed in the IR. 
One kind of model is the slice of AdS stabilized by the Goldberger--Wise mechanism. This produces a non-negligible backreaction of the metric near the IR brane. Another class of model are those where the metric develops a naked curvature singularity in the IR---the soft-wall models, see \textit{e.g.} \cite{Karch:2006pv,Gursoy:2007cb,Gursoy:2007er, Batell:2008zm, Cabrer:2009we,vonGersdorff:2010ht} for some points of entry in the literature. Such models are typically asymptotically AdS towards the UV brane, with the IR deformation becoming relevant near the IR brane/singularity.

One can apply the reasoning of Section~\ref{se:regimes}  to these models. By dimensional analysis, there is some typical scale $\hat\mu$ associated with the IR region. A transition scale $\Lambda \hat\mu/k$ thus also exists, above which the IR region should drop from the correlation functions if the EFT is to remain under control. 

More quantitatively, one can integrate out the IR region and encapsulate it into an effective IR brane with non-trivial form factors localized on it \cite{vonGersdorff:2010ht}. This holographic projection of the IR region demonstrates that the two regimes can indeed be meaningfully separated. The effective IR brane contains the details of the model-dependent KK regime. Since the bulk is pure AdS, our results from Sections~\ref{se:AdS}-\ref{se:picture}  apply. 
This immediately shows  that at high enough $p$, the effective IR brane leaves the theory, leaving thus a (quasi-)AdS continuum regime like the one described in this paper. Conversely, when decreasing $p$, the deviation from AdS gradually emerges from the viewpoint of  a UV-brane observer.

\subsection{Holographic Dark Sector}

The soft bomb suppression rate has phenomenological implications for theories where a dark (or hidden) sector is confined to the IR brane and the Standard Model is confined to the UV brane,  as recently proposed in~\cite{Brax:2019koq}.
Suppose, for concreteness, that the decay chain ends in stable IR brane particles that could naturally be identified with dark matter.

A standard way to search for dark matter at colliders is to look for missing energy signatures. In our holographic dark sector scenario, the suppression of the cascade decay rate in the continuum regime implies that the missing energy spectrum should vanish around the  $\tilde \Lambda$ scale.
This characteristic of the holographic dark sector framework is completely distinct from standard 4D dark sectors.

Another standard constraint on dark sectors with light states that couple to the Standard Model is stellar cooling from the emission of dark states. 
In the holographic dark sector scenario, stars emit KK modes with narrow widths when the temperature of the star is roughly between $\mu$ and $\tilde \Lambda$.
In contrast, if the star is hotter than $\tilde \Lambda$, the  center-of-mass energy for dark state production is typically in the continuum regime. 
One may then expect that the anomalous cooling rates are then exponentially suppressed within the AdS model. One must be cautious with this intuition, however, as the finite-temperature system may be better described by an AdS--Schwarzschild geometry~\cite{Creminelli:2001th}. The phenomenology of this situation may lead to new possibilities to get around the stellar cooling bounds that constrain standard dark sectors.

These effects are very interesting from a phenomenological viewpoint: they may alleviate experimental constraints and change the experimental complementarity of dark matter searches. The direct detection signatures of this type of framework are studied in~\cite{Katz:2015zba}, which also discusses some qualitative differences of timelike correlation functions in models of near-conformal sectors.
 We explore these effects in upcoming phenomenological studies. 

\section{AdS/CFT}\label{se:adscft}

This paper focused primarily on the physics of 5D Anti-de Sitter spacetime. In this section we connect our analysis to the properties of the dual  gauge theory by the AdS/CFT correspondence. 
First we briefly discuss consistency of our soft bomb picture with the one obtained in the CFT literature. We will then show how  dimensional analysis (see Section\,\ref{se:NDA}) applied to the holographic action naturally relates to the dual large-$N$ expansion. Finally 
 we study the transition scale in the dual low-energy EFT of glueballs. 

\subsection{CFT Soft Bombs }
\label{se:CFT_softbomb}

There is strong evidence that gauge theories with large 't\,Hooft coupling exhibit vastly different behaviour than weakly-coupled gauge theories, see {e.g.}~\cite{Polchinski:2002jw}.\,\footnote{In the gauge theory context, the strongly-coupled analog of jets have sometimes been referred to as ``spherical events" or ``jets at strong coupling" instead of ``soft bombs'' as done here. }

In \cite{Hatta:2008tn}, the fragmentation of a jet at large 't\,Hooft coupling was qualitatively studied using properties of spacelike and timelike anomalous dimensions. 
The jet is assumed to be created from  well-defined asymptotic states such as in $e^+ e^-$ annihilation.  
In our AdS dual this is realized using asymptotic states localized on the UV brane. 
The jet evolves and ends at some infrared scale $\Lambda_{\rm IR}$ at which the parton momenta are measured. 
In our AdS dual this $\Lambda_{\rm IR}$ corresponds to the infrared  scale $\tilde \Lambda$ that we have determined in Section\,\ref{se:regimes}. 
 
Ref.\,\cite{Hatta:2008tn} finds that parton splitting tends to be democratic because there is no reason for soft or collinear phase space configurations to be preferred---all partons tend to have minimum momentum $p_f\sim \Lambda_{\rm IR}$. Hence, cascades give rise to spherical events with a large number of low-momentum final states. 
This matches our explicit {AdS} calculation in {Section~\ref{se:shape}}. The total {number} of offspring is found to be $n \sim P/\Lambda_{\rm IR}$, which corresponds to {\eqref{eq:mult}}, {with} $\Lambda_{\rm IR}\sim \tilde \Lambda$. 
 
We conclude that the shape of {an} AdS soft bomb event is consistent with  findings on the CFT side.

\subsection{Dimensional Analysis and Large \texorpdfstring{$N$}{N} }\label{se:ndaadscft}

In Section\,\ref{se:matter} we have shown that 4D KK mode interactions are naturally suppressed by powers of $(\ell_5 k/ \ell_4\Lambda)$. Here we show that this suppression corresponds to the large $N$ suppression in the dual CFT.
To see this correspondence, instead of KK modes, we must consider the 5D theory in AdS using an appropriate variable---the value of the bulk field on the UV brane 
 \begin{align}
 \hat \Phi_0(x)\equiv \hat \Phi(X)\Big|_{\rm UV\,brane} \, .
 \end{align}
$\Phi$ is the dimensionless bulk field in ~\eqref{eq:SD_NDA}.
The bulk field in the action is rewritten as $\hat \Phi=\hat \Phi_0 K$, where $K$ is the classical field profile sourced by $\hat \Phi_0$.
In terms of this holographic variable, the partition function~\eqref{eq:Zfull} takes the form $\int {\cal D} \hat \Phi_0 \exp \left( i S_5[\hat \Phi_0 K]\right)$, where $S_5$ is the 5D action for which the 5D NDA in~\eqref{eq:SD_NDA} applies. 

The leading term of the effective action in the semiclassical expansion is the classical holographic action
\begin{align}
\Gamma_{\rm hol} = \frac{\Lambda^5}{\ell_5} \int d^4 x  {\cal L}_{\rm hol}\left[\hat \Phi_0, \partial/\Lambda\right]+\cdots\,
\label{eq:SD_NDA_hol},
\end{align}
where the ellipses represent quantum terms that are irrelevant for our discussion. 
The Lagrangian  ${\mathcal L}_\textnormal{hol}$  has dimension $-1$. To recover a 4D NDA formulation as in \eqref{eq:SD_NDA}, we need to introduce a dimensionless Lagrangian. 
From explicit calculation (see {e.g.}~\cite{Witten:1998qj, Ponton:2012bi}), the quadratic part of  ${\mathcal L}_\textnormal{hol}$, $\frac{1}{2}\hat\Phi_0\Pi[\partial^2] \hat\Phi_0$, is proportional to the inverse of $\Delta_q(z_0,z_0)$ and contains an analytic part representing a 4D mode. Schematically, it is
\begin{align}
\Pi[\partial^2]\sim -\frac{1}{k}\frac{\partial^2+m_0^2}{\Lambda^2} + \ldots
\label{eq:Pi}
\end{align}
up to an $\mathcal O(1)$ coefficient. In the language of AdS/CFT, this is the kinetic term of the 4D source probing the CFT. 
The exact expression can be read directly from the propagator~\eqref{eq:propaII} and is not needed here. 

We introduce the dimensionless Lagrangian  $\frac{1}{k}\hat{\cal L}_{\rm hol}={\cal L}_{\rm hol}$, such that the dimensionless source described in~\eqref{eq:Pi}  is canonically normalized.  
The action now can be rearranged as
\begin{align}
\Gamma_{\rm hol} = \left(\frac{\ell_4 \Lambda }{ \ell_5 k} \right)  \frac{\Lambda^4}{\ell_4} \int d^4 x \hat {\cal L}_{\rm hol}\left[\hat \Phi_0, \partial/\Lambda\right]+\cdots \,, \label{eq:Gammahol}
\end{align}
where we explicitly write the $\Lambda^4/\ell_4$ factor appear in accordance with 4D NDA. The factor in parenthesis is the same suppression as obtained in Section~\ref{se:matter}. 
From \eqref{eq:SD_NDA_hol} it is clear that this factor systematically appears alongside $\hbar$ in the semiclassical expansion of the holographic action.

We may now perform dimensional analysis on the canonically normalized holographic variable, 
\begin{align}
\Phi_0= \left(\frac{\ell_4 \Lambda }{ \ell_5 k} \right)^{1/2}  \frac{\Lambda}{\ell_4^{1/2}}  \hat \Phi_0 \, .
\end{align}
Functional derivatives with respect to $\Phi_0$ are suppressed  as 
\begin{align}
\frac{\delta^n\Gamma_{\rm hol}}{\delta \Phi_0(x_1)\delta \Phi_0(x_2)\cdots} \propto \left(\frac{ \ell_5 k}{\ell_4 \Lambda } \right)^{n/2-1}\, \label{eq:AdS_corr}
\end{align}
at leading order. 
Hence by applying dimensional analysis at 5D and 4D levels in the holographic action, we have shown that a small parameter ($\sqrt{{\ell_5 k}/{\ell_4 \Lambda}}$)  systematically suppressing the interactions  and controlled by the AdS curvature appears.

The AdS/CFT correspondence dictates that the above quantity reproduces the connected $n$-point functions of a conformal gauge theory with adjoint fields and large $N$. 
The main contribution to the correlator at large $N$ is  suppressed as~\cite{Witten:1979kh,Nastase:2007kj} 
\begin{align}
\langle {\cal O} {\cal O}\ldots \rangle_{\rm con} \propto \frac{1}{N^{n-2}}\, 
\label{eq:CFT_corr}
\end{align}
with canonical normalization such that the 2pt function does not scale with $N$.
Comparing the AdS expression
\eqref{eq:AdS_corr} and the CFT expression~\eqref{eq:CFT_corr}, we  see that the suppression factor in AdS corresponds to  the $1/N^2$ suppression of the CFT, 
\begin{align}
\frac{ \ell_5 k}{\ell_4 \Lambda }\sim \frac{1}{N^2}\,. \label{eq:AdS_N}
\end{align}
We thus obtain a precise, field-theoretical version of the correspondence between the $1/N$ expansion in the CFT and the parameters of the AdS effective field theory. 
At fixed AdS curvature $k$, and i.e.~fixed 't\,Hooft coupling, the $N\rightarrow \infty$ limit corresponds to the $\Lambda\rightarrow \infty$ limit. This sets all interactions to zero and therefore produces a free 5D theory.
The relation~\eqref{eq:AdS_N}, when put in the 
holographic action~\eqref{eq:Gammahol}, gives
$\Gamma_{\rm hol}=N^2 \frac{\Lambda^4}{{\ell_4}} \int d^4 x \hat{\cal L}_{\rm hol}$. The $N^2$ factor accompanying $\hbar$ in this action is  a hallmark feature of AdS/CFT\,\cite{Aharony:1999ti}.

\subsection{{Dual} Interpretation of Transition Scale }

\label{se:CFT_glueballs}

In this section we consider the {dual gauge theory} interpretation of the transition scale using $\Lambda/k\sim N^2 \ell_5/\ell_4$ as established in \eqref{eq:AdS_N}. In the following discussion, we estimate $\ell_5/\ell_4 \sim \pi$.
Interactions vanish in the $\Lambda\rightarrow \infty $ limit. In this limit, the AdS theory thus contains an infinite tower of free, stable KK modes.\,\footnote{{A confined gauge theory produces a spectrum with particles of arbitrary spin. In the QFT approach to AdS/QCD, glueballs of a given spin corresponds to a given field on the AdS side, see \textit{e.g.}~\cite{Karch:2006pv}. Focusing on a bulk scalar amounts to focusing on the sector of scalar glueballs.}
}
This is the AdS manifestation of the infinite tower of stable glueballs when $N\rightarrow \infty $.

For finite $N$, the transition scale is
\begin{align}\tilde{\Lambda}  \sim N^2 \pi \mu \,. \label{eq:Lambdatilde}
\end{align}
The scale controlling the mass  of the KK modes, $\pi \mu$ appears. The KK masses grow linearly, hence the transition is reached around the mass of the $N^{2}\textnormal{-th}$ KK mode. 

The $\tilde \Lambda$ scale would be the cutoff of the glueball EFT. Does the value \eqref{eq:Lambdatilde} make sense from the gauge theory side? 
Recall that the  large-$N$  theory contains, in principle, many glueballs at low energy. It is thus described by an  EFT containing many species.
The interactions between glueballs are set by the $\tilde \Lambda$ scale and suppressed by powers of $1/N$.  In the loop diagrams, such suppression is compensated by the multiplicity of glueballs.
For $N^2$ glueballs, the cutoff of the EFT becomes $\tilde \Lambda$.
This feature can be  seen by using 4D NDA applied to the glueball theory with arbitrary number of species $N_s$ and $D=4$.
The prefactor of \eqref{eq:SD_NDA}  is
\begin{align}\frac{N_s\tilde\Lambda^4}{N^2 \ell_4}\,,
\end{align}
which indicates strong coupling  when the number of glueballs $N_s$ is of order $ N^2$. 
This paints a consistent picture: the $N^2$ modes of the KK regime in AdS match the $N^2$ glueballs of the gauge theory.

\begin{figure}[t]
\center
	\includegraphics[width=0.60\linewidth]{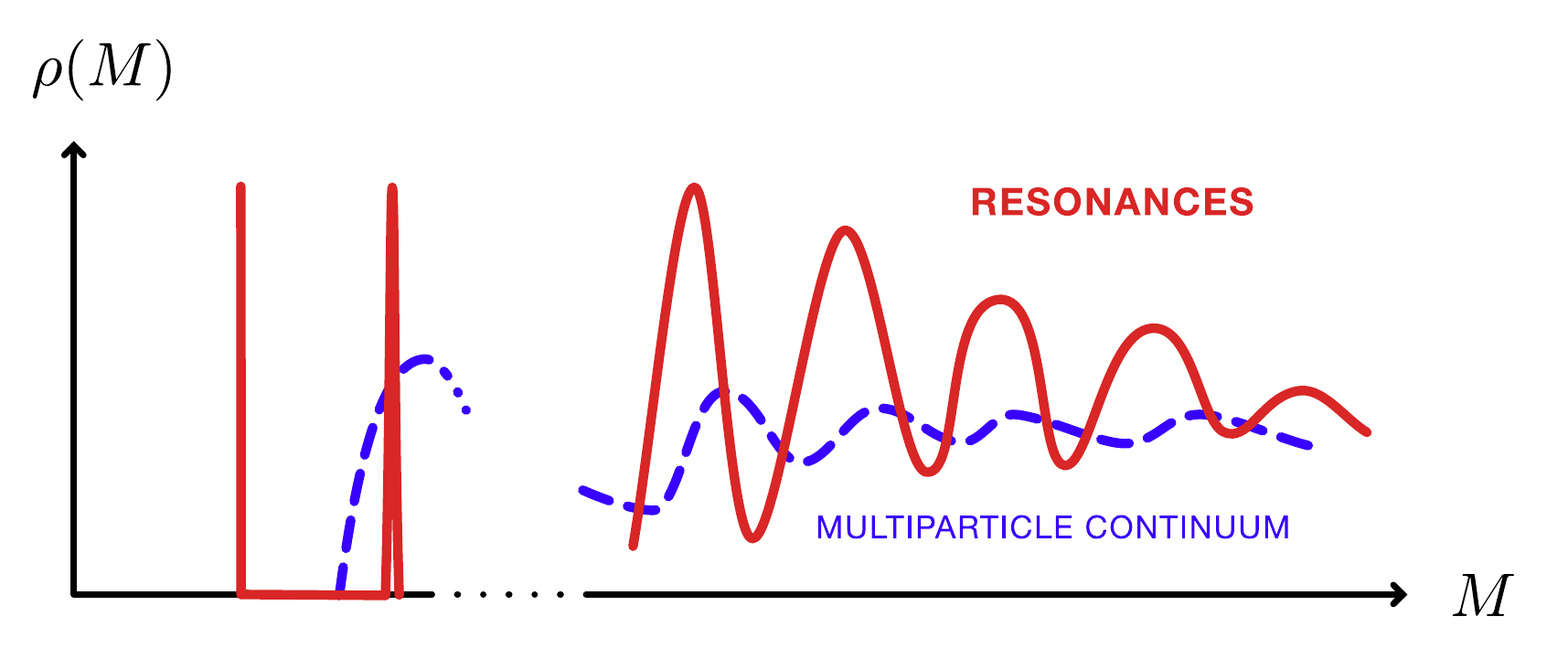}
		\caption{
		Schematic spectral density of the two-point correlator of the large-$N$ glueball EFT. 
		The solid line shows the glueball resonances merging into a continuum when approaching the cutoff of the EFT. 
		The merging of resonances that we describe is distinct from the multiparticle continuum, which we show schematically for completeness. 
		\label{fig:balls} 
		}
\end{figure}

These considerations are only about the number of species and do not tell us about glueball masses. 
However we also know that an infinite tower of glueballs is needed to reproduce the logarithmic momentum dependence of the  correlator between gauge currents~\cite{Witten:1979kh}. 
At finite $N$,  the width over mass ratio of the $n^\textnormal{th}$ glueball is expected to grow as $\Gamma_n/m_n \sim n/N^2$.
The $(1/N)^2$ factor comes from the $1/N$ suppressed cubic vertices, and the $n$ factor comes from the number of accessible decay channels into lighter glueballs. 
Hence the glueballs tend to become broad at $n\sim N^2$, which  signals the transition to a  continuum.
Since there is a  tower of glueballs, the cutoff of the glueball EFT has to be around the mass of the $N^2\textnormal{-th}$ glueball, \textit{i.e.}~$\tilde \Lambda \sim m_{N^2}$.
This matches the picture obtained on  the AdS side in \eqref{eq:Lambdatilde}, where the $N^2\textnormal{-th}$ KK mode is indeed of order of $\tilde\Lambda$.
{Notice this reasoning relies on state-counting and only requires enough of a hierarchy between masses for the decays to occur. This is a very mild condition. 
In this work the mass distribution obtained from the AdS side  is $m_n\sim n$, but  the same line of reasoning  would apply to \textit{e.g.} a Regge-like spectrum $m_n\sim \sqrt{n}$ }\,.\,\footnote{For completeness 
we notice that at energies approaching the cutoff, $\tilde\Lambda$,
contributions from the multiparticle continuum should become sizeable. This is expected since, by definition, at the EFT cutoff the contributions from all loop orders become same order of magnitude---the EFT becomes strongly coupled. 
The multiparticle contributions are suppressed by additional $1/N$ factors and by loop factors $1/\ell_4$ which define the perturbative expansion in the EFT. Near the cutoff, the $1/N$ suppression is compensated by the multiplicity of states in the loop(s). 
Hence near the cutoff, one both expects that the resonances merge into a tree-level continuum and that loop-level continua become of same order as the tree-level one. 
This is included in Figure~\ref{fig:balls}. }

The spectral density of glueballs obained from the above considerations  is summarized in Fig.\,\ref{fig:balls}.

 \section{Conclusion}
\label{se:con}

We revisit the behaviour of an effective theory of interacting matter fields in a slice of AdS$_5$.
We work in Poincar\'e position-momentum space---the AdS Poincar\'e patch Fourier-transformed along Minkowski slices. 

We study new features induced by bulk interactions for timelike four-momenta. These correspond to including the leading $1/N^2$ effects in the strongly coupled dual theory.
We show using dimensional analysis that there is a transition scale, $\tilde \Lambda$, above which bulk propagators lose contact with the IR brane because the latter falls beyond the domain of validity of the effective theory.  The scale separates the Kaluza--Klein and continuum regimes of the bulk propagator.  The continuum regime would be absent if interactions were not taken into account. Conversely the continuum regime is the only one present in the limit of strong interactions.

For timelike momenta the transition between the KK and continuum regimes occurs because the propagator is dressed by bulk interactions, a leading $1/N^2$ effect. This induces an exponential suppression of the propagator in the region where the EFT would become invalid.
This censorship property was qualitatively predicted in \cite{ArkaniHamed:2000ds}.
Our treatment invokes approximations to loop integrals; more details of opacity in AdS may be better elucidated with future formal developments.

In the CFT dual, the existence of the transition scale corresponds to the fact that the effective theory of glueballs cannot contain infinitely many species. It becomes strongly coupled if more than approximately $ N^2$ glueballs are included in the spectrum.  Beyond the transition scale, a gauge theory with no mass gap should appear. This is indeed what we demonstrate in the AdS theory.

For timelike bulk propagators, the IR brane is effectively absent when $p> \tilde \Lambda$. However, cascade decays could allow correlators with energy beyond  $\tilde \Lambda$ to be sensitive to the IR brane because the momentum is split between many offspring states. We therefore study cascade decays to better understand the notion of IR brane emergence.
We focus on a scalar with a bulk cubic interaction and investigate the squared matrix element integrated over final states that are the main ingredients of observable event rates.
In the continuum regime, there exists a recursion relation between cascades of different branching depth, which we use to estimate  the rate for arbitrarily deep cascade.

We have checked that contributions from other effects are subleading.
These include direct decays into an IR brane localized state or into light KK modes via a tower of off-shell KK modes. We found that the contribution from the region in which the propagator is exponentially suppressed may be of the same order, but that this does not change our conclusions.

The cascade decay calculation provides a picture of soft bombs in the continuum regime of AdS. 
We find that the shape of the cascade events tend to be soft and spherical in the 4D Minkowski slices.
This is because the branchings tend to be near-threshold with momentum evenly split between the offspring, which matches previous results for the CFT dual. Along the  fifth  dimension, the decays tend to occur near the region $z\sim 1/p$ where $p$ is the parent four-momentum. Therefore the soft bomb diagram grows in the Minkowski direction and slowly progresses towards the IR.  Once the typical momentum of the offspring reaches $\tilde \Lambda$, the soft bomb enters the KK regime.

While there is no diagrammatic change between the KK and continuum regimes, the crucial change occurs in the behaviour of the propagators. In the KK regime, the narrow width approximation applies, such that amplitudes giving the soft bomb rate can effectively be cut. In the continuum regime the propagators do not have poles and the event cannot be cut before reaching the KK regime. 
The phase space factor associated with each of the final states accumulate and the soft bomb rate in the continuum regime acquires an exponential suppression. 
It follows that the continuum regime can be described by a high-energy effective theory with no IR brane. In other words, the operators on the IR brane effectively emerge at the energy scale $E\sim \tilde \Lambda$, 
\textit{i.e.} schematically $
E[J]\big|_{p\gg \tilde \Lambda } \approx E[J]_{\rm no\, IR\, brane}
$ in terms of generating functionals of correlators.
We expect that the same conclusions qualitatively apply to asymptotically AdS backgrounds with a metric deformation in the IR region, such as soft-wall models.

These features can lead to new possibilities for physics beyond the Standard Model, as already pointed out in \cite{Brax:2019koq}.
In particular holographic dark sector scenarios may have bulk fields that mediate interactions between a UV-brane localized  Standard Model and IR brane dark states that are emergent.
This implies that a light dark particle can be invisible at high energy experiments. For instance, bounds from stellar cooling or missing energy searches may be alleviated if the dark particles are light enough. The many phenomenological consequences of an emergent dark sector require further studies.

 \section*{Acknowledgements}

We thank  K.~Agashe, S.~Belayev, D. Buarque,  Z.~Chako, H.~Davoudiasl, J.~Hubisz, M.~Luty,  R.~Sundrum, and G.~von Gersdorff for useful discussions. 
A.C.~is supported by the National Science Foundation Graduate Research Fellowship Program under Grant No.~1840991.
S.F.~is supported by the S\~ao Paulo Research Foundation (FAPESP) under grants \#2011/11973, \#2014/21477-2 and \#2018/11721-4, and funded in part by the Gordon and Betty Moore Foundation through a Fundamental Physics Innovation Visitor Award (Grant GBMF6210).
P.T.~is supported by~{de-sc}/0008541. P.T.~thanks the Aspen Center for Physics ({NSF} grant \#1066293) and the Kavli Institute for Theoretical Physics ({NSF} grant {PHY}-1748958) for their hospitality while part of this work was completed.


\bibliographystyle{utcaps} 	
\bibliography{SoftBomb}

\providecommand{\href}[2]{#2}\begingroup\raggedright\begin{thebibliography}{10}

\bibitem{Maldacena:1997re}
J.~M. Maldacena, ``{The Large N limit of superconformal field theories and
  supergravity},'' \href{http://dx.doi.org/10.1023/A:1026654312961,
  10.4310/ATMP.1998.v2.n2.a1}{{\em Int. J. Theor. Phys.} {\bfseries 38} (1999)
  1113--1133}, \href{http://arxiv.org/abs/hep-th/9711200}{{\ttfamily
  arXiv:hep-th/9711200 [hep-th]}}.
[Adv. Theor. Math. Phys.2,231(1998)].

\bibitem{Gubser:1998bc}
S.~S. Gubser, I.~R. Klebanov, and A.~M. Polyakov, ``{Gauge theory correlators
  from noncritical string theory},''
  \href{http://dx.doi.org/10.1016/S0370-2693(98)00377-3}{{\em Phys. Lett.}
  {\bfseries B428} (1998) 105--114},
\href{http://arxiv.org/abs/hep-th/9802109}{{\ttfamily arXiv:hep-th/9802109
  [hep-th]}}.

\bibitem{Witten:1998qj}
E.~Witten, ``{Anti-de Sitter space and holography},''
  \href{http://dx.doi.org/10.4310/ATMP.1998.v2.n2.a2}{{\em Adv. Theor. Math.
  Phys.} {\bfseries 2} (1998) 253--291},
\href{http://arxiv.org/abs/hep-th/9802150}{{\ttfamily arXiv:hep-th/9802150
  [hep-th]}}.

\bibitem{Aharony:1999ti}
O.~Aharony, S.~S. Gubser, J.~M. Maldacena, H.~Ooguri, and Y.~Oz, ``{Large N
  field theories, string theory and gravity},''
  \href{http://dx.doi.org/10.1016/S0370-1573(99)00083-6}{{\em Phys. Rept.}
  {\bfseries 323} (2000) 183--386},
\href{http://arxiv.org/abs/hep-th/9905111}{{\ttfamily arXiv:hep-th/9905111
  [hep-th]}}.

\bibitem{Heemskerk:2009pn}
I.~Heemskerk, J.~Penedones, J.~Polchinski, and J.~Sully, ``{Holography from
  Conformal Field Theory},''
  \href{http://dx.doi.org/10.1088/1126-6708/2009/10/079}{{\em JHEP} {\bfseries
  10} (2009) 079},
\href{http://arxiv.org/abs/0907.0151}{{\ttfamily arXiv:0907.0151 [hep-th]}}.

\bibitem{Heemskerk:2010ty}
I.~Heemskerk and J.~Sully, ``{More Holography from Conformal Field Theory},''
  \href{http://dx.doi.org/10.1007/JHEP09(2010)099}{{\em JHEP} {\bfseries 09}
  (2010) 099},
\href{http://arxiv.org/abs/1006.0976}{{\ttfamily arXiv:1006.0976 [hep-th]}}.

\bibitem{Fitzpatrick:2010zm}
A.~L. Fitzpatrick, E.~Katz, D.~Poland, and D.~Simmons-Duffin, ``{Effective
  Conformal Theory and the Flat-Space Limit of AdS},''
  \href{http://dx.doi.org/10.1007/JHEP07(2011)023}{{\em JHEP} {\bfseries 07}
  (2011) 023},
\href{http://arxiv.org/abs/1007.2412}{{\ttfamily arXiv:1007.2412 [hep-th]}}.

\bibitem{Sundrum:2011ic}
R.~Sundrum, ``{From Fixed Points to the Fifth Dimension},''
  \href{http://dx.doi.org/10.1103/PhysRevD.86.085025}{{\em Phys. Rev.}
  {\bfseries D86} (2012) 085025},
\href{http://arxiv.org/abs/1106.4501}{{\ttfamily arXiv:1106.4501 [hep-th]}}.

\bibitem{ElShowk:2011ag}
S.~El-Showk and K.~Papadodimas, ``{Emergent Spacetime and Holographic CFTs},''
  \href{http://dx.doi.org/10.1007/JHEP10(2012)106}{{\em JHEP} {\bfseries 10}
  (2012) 106},
\href{http://arxiv.org/abs/1101.4163}{{\ttfamily arXiv:1101.4163 [hep-th]}}.

\bibitem{Fitzpatrick:2012cg}
A.~L. Fitzpatrick and J.~Kaplan, ``{AdS Field Theory from Conformal Field
  Theory},'' \href{http://dx.doi.org/10.1007/JHEP02(2013)054}{{\em JHEP}
  {\bfseries 02} (2013) 054},
\href{http://arxiv.org/abs/1208.0337}{{\ttfamily arXiv:1208.0337 [hep-th]}}.

\bibitem{Karch:2006pv}
A.~Karch, E.~Katz, D.~T. Son, and M.~A. Stephanov, ``{Linear confinement and
  AdS/QCD},'' \href{http://dx.doi.org/10.1103/PhysRevD.74.015005}{{\em Phys.
  Rev.} {\bfseries D74} (2006) 015005},
\href{http://arxiv.org/abs/hep-ph/0602229}{{\ttfamily arXiv:hep-ph/0602229
  [hep-ph]}}.

\bibitem{Gursoy:2007cb}
U.~Gursoy and E.~Kiritsis, ``{Exploring improved holographic theories for QCD:
  Part I},'' \href{http://dx.doi.org/10.1088/1126-6708/2008/02/032}{{\em JHEP}
  {\bfseries 02} (2008) 032},
\href{http://arxiv.org/abs/0707.1324}{{\ttfamily arXiv:0707.1324 [hep-th]}}.

\bibitem{Gursoy:2007er}
U.~Gursoy, E.~Kiritsis, and F.~Nitti, ``{Exploring improved holographic
  theories for QCD: Part II},''
  \href{http://dx.doi.org/10.1088/1126-6708/2008/02/019}{{\em JHEP} {\bfseries
  02} (2008) 019},
\href{http://arxiv.org/abs/0707.1349}{{\ttfamily arXiv:0707.1349 [hep-th]}}.

\bibitem{Gubser:2008ny}
S.~S. Gubser and A.~Nellore, ``{Mimicking the QCD equation of state with a dual
  black hole},'' \href{http://dx.doi.org/10.1103/PhysRevD.78.086007}{{\em Phys.
  Rev. D} {\bfseries 78} (2008) 086007},
  \href{http://arxiv.org/abs/0804.0434}{{\ttfamily arXiv:0804.0434 [hep-th]}}.

\bibitem{Falkowski:2008fz}
A.~Falkowski and M.~Perez-Victoria, ``{Electroweak Breaking on a Soft Wall},''
  \href{http://dx.doi.org/10.1088/1126-6708/2008/12/107}{{\em JHEP} {\bfseries
  12} (2008) 107},
\href{http://arxiv.org/abs/0806.1737}{{\ttfamily arXiv:0806.1737 [hep-ph]}}.

\bibitem{Batell:2008zm}
B.~Batell and T.~Gherghetta, ``{Dynamical Soft-Wall AdS/QCD},''
  \href{http://dx.doi.org/10.1103/PhysRevD.78.026002}{{\em Phys. Rev.}
  {\bfseries D78} (2008) 026002},
\href{http://arxiv.org/abs/0801.4383}{{\ttfamily arXiv:0801.4383 [hep-ph]}}.

\bibitem{Batell:2008me}
B.~Batell, T.~Gherghetta, and D.~Sword, ``{The Soft-Wall Standard Model},''
  \href{http://dx.doi.org/10.1103/PhysRevD.78.116011}{{\em Phys. Rev.}
  {\bfseries D78} (2008) 116011},
\href{http://arxiv.org/abs/0808.3977}{{\ttfamily arXiv:0808.3977 [hep-ph]}}.

\bibitem{Cabrer:2009we}
J.~A. Cabrer, G.~von Gersdorff, and M.~Quiros, ``{Soft-Wall Stabilization},''
  \href{http://dx.doi.org/10.1088/1367-2630/12/7/075012}{{\em New J. Phys.}
  {\bfseries 12} (2010) 075012},
\href{http://arxiv.org/abs/0907.5361}{{\ttfamily arXiv:0907.5361 [hep-ph]}}.

\bibitem{vonGersdorff:2010ht}
G.~von Gersdorff, ``{From Soft Walls to Infrared Branes},''
  \href{http://dx.doi.org/10.1103/PhysRevD.82.086010}{{\em Phys. Rev.}
  {\bfseries D82} (2010) 086010},
\href{http://arxiv.org/abs/1005.5134}{{\ttfamily arXiv:1005.5134 [hep-ph]}}.

\bibitem{Cabrer:2011fb}
J.~A. Cabrer, G.~von Gersdorff, and M.~Quiros, ``{Suppressing Electroweak
  Precision Observables in 5D Warped Models},''
  \href{http://dx.doi.org/10.1007/JHEP05(2011)083}{{\em JHEP} {\bfseries 05}
  (2011) 083},
\href{http://arxiv.org/abs/1103.1388}{{\ttfamily arXiv:1103.1388 [hep-ph]}}.

\bibitem{Megias:2019vdb}
E.~Meg{\'\i}as and M.~Quir{\'o}s, ``{Gapped Continuum Kaluza-Klein spectrum},''
  \href{http://dx.doi.org/10.1007/JHEP08(2019)166}{{\em JHEP} {\bfseries 08}
  (2019) 166},
\href{http://arxiv.org/abs/1905.07364}{{\ttfamily arXiv:1905.07364 [hep-ph]}}.

\bibitem{Hofman:2008ar}
D.~M. Hofman and J.~Maldacena, ``{Conformal collider physics: Energy and charge
  correlations},'' \href{http://dx.doi.org/10.1088/1126-6708/2008/05/012}{{\em
  JHEP} {\bfseries 05} (2008) 012},
\href{http://arxiv.org/abs/0803.1467}{{\ttfamily arXiv:0803.1467 [hep-th]}}.

\bibitem{Chesler:2008wd}
P.~M. Chesler, K.~Jensen, and A.~Karch, ``{Jets in strongly-coupled N = 4 super
  Yang-Mills theory},''
  \href{http://dx.doi.org/10.1103/PhysRevD.79.025021}{{\em Phys. Rev.}
  {\bfseries D79} (2009) 025021},
\href{http://arxiv.org/abs/0804.3110}{{\ttfamily arXiv:0804.3110 [hep-th]}}.

\bibitem{Hatta:2007cs}
Y.~Hatta, E.~Iancu, and A.~H. Mueller, ``{Deep inelastic scattering off a N=4
  SYM plasma at strong coupling},''
  \href{http://dx.doi.org/10.1088/1126-6708/2008/01/063}{{\em JHEP} {\bfseries
  01} (2008) 063},
\href{http://arxiv.org/abs/0710.5297}{{\ttfamily arXiv:0710.5297 [hep-th]}}.

\bibitem{Hatta:2008tx}
Y.~Hatta, E.~Iancu, and A.~H. Mueller, ``{Jet evolution in the N=4 SYM plasma
  at strong coupling},''
  \href{http://dx.doi.org/10.1088/1126-6708/2008/05/037}{{\em JHEP} {\bfseries
  05} (2008) 037},
\href{http://arxiv.org/abs/0803.2481}{{\ttfamily arXiv:0803.2481 [hep-th]}}.

\bibitem{Hatta:2008qx}
Y.~Hatta and T.~Matsuo, ``{Thermal hadron spectrum in e+e- annihilation from
  gauge/string duality},''
  \href{http://dx.doi.org/10.1103/PhysRevLett.102.062001}{{\em Phys. Rev.
  Lett.} {\bfseries 102} (2009) 062001},
\href{http://arxiv.org/abs/0807.0098}{{\ttfamily arXiv:0807.0098 [hep-ph]}}.

\bibitem{Hatta:2008tn}
Y.~Hatta and T.~Matsuo, ``{Jet fragmentation and gauge/string duality},''
  \href{http://dx.doi.org/10.1016/j.physletb.2008.10.043}{{\em Phys. Lett.}
  {\bfseries B670} (2008) 150--153},
\href{http://arxiv.org/abs/0804.4733}{{\ttfamily arXiv:0804.4733 [hep-th]}}.

\bibitem{Knapen:2016hky}
S.~Knapen, S.~Pagan~Griso, M.~Papucci, and D.~J. Robinson, ``{Triggering Soft
  Bombs at the LHC},'' \href{http://dx.doi.org/10.1007/JHEP08(2017)076}{{\em
  JHEP} {\bfseries 08} (2017) 076},
\href{http://arxiv.org/abs/1612.00850}{{\ttfamily arXiv:1612.00850 [hep-ph]}}.

\bibitem{Csaki:2008dt}
C.~Csaki, M.~Reece, and J.~Terning, ``{The AdS/QCD Correspondence: Still
  Undelivered},'' \href{http://dx.doi.org/10.1088/1126-6708/2009/05/067}{{\em
  JHEP} {\bfseries 05} (2009) 067},
\href{http://arxiv.org/abs/0811.3001}{{\ttfamily arXiv:0811.3001 [hep-ph]}}.

\bibitem{ArkaniHamed:2000ds}
N.~Arkani-Hamed, M.~Porrati, and L.~Randall, ``{Holography and
  phenomenology},'' \href{http://dx.doi.org/10.1088/1126-6708/2001/08/017}{{\em
  JHEP} {\bfseries 08} (2001) 017},
\href{http://arxiv.org/abs/hep-th/0012148}{{\ttfamily arXiv:hep-th/0012148
  [hep-th]}}.

\bibitem{Goldberger:2002cz}
W.~D. Goldberger and I.~Z. Rothstein, ``{High-energy field theory in truncated
  AdS backgrounds},''
  \href{http://dx.doi.org/10.1103/PhysRevLett.89.131601}{{\em Phys. Rev. Lett.}
  {\bfseries 89} (2002) 131601},
\href{http://arxiv.org/abs/hep-th/0204160}{{\ttfamily arXiv:hep-th/0204160
  [hep-th]}}.

\bibitem{Fichet:2019hkg}
S.~Fichet, ``{Opacity and effective field theory in anti--de Sitter
  backgrounds},'' \href{http://dx.doi.org/10.1103/PhysRevD.100.095002}{{\em
  Phys. Rev.} {\bfseries D100} no.~9, (2019) 095002},
\href{http://arxiv.org/abs/1905.05779}{{\ttfamily arXiv:1905.05779 [hep-th]}}.

\bibitem{Brax:2019koq}
P.~Brax, S.~Fichet, and P.~Tanedo, ``{The Warped Dark Sector},''
  \href{http://dx.doi.org/10.1016/j.physletb.2019.135012}{{\em Phys. Lett.}
  {\bfseries B798} (2019) 135012},
\href{http://arxiv.org/abs/1906.02199}{{\ttfamily arXiv:1906.02199 [hep-ph]}}.

\bibitem{Costantino:2019ixl}
A.~Costantino, S.~Fichet, and P.~Tanedo, ``{Exotic Spin-Dependent Forces from a
  Hidden Sector},''
\href{http://arxiv.org/abs/1910.02972}{{\ttfamily arXiv:1910.02972 [hep-ph]}}.

\bibitem{vonHarling:2012sz}
B.~von Harling and K.~L. McDonald, ``{Secluded Dark Matter Coupled to a Hidden
  CFT},'' \href{http://dx.doi.org/10.1007/JHEP08(2012)048}{{\em JHEP}
  {\bfseries 08} (2012) 048},
\href{http://arxiv.org/abs/1203.6646}{{\ttfamily arXiv:1203.6646 [hep-ph]}}.

\bibitem{McDonald:2012nc}
K.~L. McDonald, ``{Sommerfeld Enhancement from Multiple Mediators},''
  \href{http://dx.doi.org/10.1007/JHEP07(2012)145}{{\em JHEP} {\bfseries 07}
  (2012) 145},
\href{http://arxiv.org/abs/1203.6341}{{\ttfamily arXiv:1203.6341 [hep-ph]}}.

\bibitem{McDonald:2010fe}
K.~L. McDonald and D.~E. Morrissey, ``{Low-Energy Signals from Kinetic Mixing
  with a Warped Abelian Hidden Sector},''
  \href{http://dx.doi.org/10.1007/JHEP02(2011)087}{{\em JHEP} {\bfseries 02}
  (2011) 087},
\href{http://arxiv.org/abs/1010.5999}{{\ttfamily arXiv:1010.5999 [hep-ph]}}.

\bibitem{McDonald:2010iq}
K.~L. McDonald and D.~E. Morrissey, ``{Low-Energy Probes of a Warped Extra
  Dimension},'' \href{http://dx.doi.org/10.1007/JHEP05(2010)056}{{\em JHEP}
  {\bfseries 05} (2010) 056},
\href{http://arxiv.org/abs/1002.3361}{{\ttfamily arXiv:1002.3361 [hep-ph]}}.

\bibitem{Katz:2015zba}
A.~Katz, M.~Reece, and A.~Sajjad, ``{Continuum-mediated dark matter--baryon
  scattering},'' \href{http://dx.doi.org/10.1016/j.dark.2016.01.002}{{\em Phys.
  Dark Univ.} {\bfseries 12} (2016) 24--36},
\href{http://arxiv.org/abs/1509.03628}{{\ttfamily arXiv:1509.03628 [hep-ph]}}.

\bibitem{Strassler:2008bv}
M.~J. Strassler, ``{Why Unparticle Models with Mass Gaps are Examples of Hidden
  Valleys},''
\href{http://arxiv.org/abs/0801.0629}{{\ttfamily arXiv:0801.0629 [hep-ph]}}.

\bibitem{Freedman:1999gp}
D.~Freedman, S.~Gubser, K.~Pilch, and N.~Warner, ``{Renormalization group flows
  from holography supersymmetry and a c theorem},''
  \href{http://dx.doi.org/10.4310/ATMP.1999.v3.n2.a7}{{\em Adv. Theor. Math.
  Phys.} {\bfseries 3} (1999) 363--417},
  \href{http://arxiv.org/abs/hep-th/9904017}{{\ttfamily arXiv:hep-th/9904017}}.

\bibitem{Breitenlohner:1982jf}
P.~Breitenl{\"o}hner and D.~Z. Freedman, ``{Stability in Gauged Extended
  Supergravity},''
\href{http://dx.doi.org/10.1016/0003-4916(82)90116-6}{{\em Annals Phys.}
  {\bfseries 144} (1982) 249}.

\bibitem{Breitenlohner:1982bm}
P.~Breitenlohner and D.~Z. Freedman, ``{Positive Energy in anti-De Sitter
  Backgrounds and Gauged Extended Supergravity},''
\href{http://dx.doi.org/10.1016/0370-2693(82)90643-8}{{\em Phys. Lett.}
  {\bfseries 115B} (1982) 197--201}.

\bibitem{Ponton:2012bi}
E.~Ponton, \href{http://dx.doi.org/10.1142/9789814390163_0007}{``{TASI 2011:
  Four Lectures on TeV Scale Extra Dimensions},''} in {\em {The Dark Secrets of
  the Terascale: Proceedings, TASI 2011, Boulder, Colorado, USA, Jun 6 - Jul
  11, 2011}}, pp.~283--374.
\newblock 2013.
\newblock
\href{http://arxiv.org/abs/1207.3827}{{\ttfamily arXiv:1207.3827 [hep-ph]}}.
\newblock

\bibitem{Fichet:2019owx}
S.~Fichet, ``{Braneworld Effective Field Theories-Holography, Consistency and
  Conformal Effects},''
\href{http://arxiv.org/abs/1912.12316}{{\ttfamily arXiv:1912.12316 [hep-th]}}.

\bibitem{Manohar:1983md}
A.~Manohar and H.~Georgi, ``{Chiral Quarks and the Nonrelativistic Quark
  Model},''
\href{http://dx.doi.org/10.1016/0550-3213(84)90231-1}{{\em Nucl. Phys.}
  {\bfseries B234} (1984) 189--212}.

\bibitem{Georgi:1986kr}
H.~Georgi and L.~Randall, ``{Flavor Conserving CP Violation in Invisible Axion
  Models},''
\href{http://dx.doi.org/10.1016/0550-3213(86)90022-2}{{\em Nucl. Phys.}
  {\bfseries B276} (1986) 241--252}.

\bibitem{Georgi:1992dw}
H.~Georgi, ``{Generalized Dimensional Analysis},''
  \href{http://dx.doi.org/10.1016/0370-2693(93)91728-6}{{\em Phys. Lett.}
  {\bfseries B298} (1993) 187--189},
\href{http://arxiv.org/abs/hep-ph/9207278}{{\ttfamily arXiv:hep-ph/9207278
  [hep-ph]}}.

\bibitem{Chacko:1999hg}
Z.~Chacko, M.~A. Luty, and E.~Ponton, ``{Massive higher dimensional gauge
  fields as messengers of supersymmetry breaking},''
  \href{http://dx.doi.org/10.1088/1126-6708/2000/07/036}{{\em JHEP} {\bfseries
  07} (2000) 036},
\href{http://arxiv.org/abs/hep-ph/9909248}{{\ttfamily arXiv:hep-ph/9909248
  [hep-ph]}}.

\bibitem{Panico:2015jxa}
G.~Panico and A.~Wulzer, ``{The Composite Nambu-Goldstone Higgs},''
  \href{http://dx.doi.org/10.1007/978-3-319-22617-0}{{\em Lect. Notes Phys.}
  {\bfseries 913} (2016) pp.1--316},
\href{http://arxiv.org/abs/1506.01961}{{\ttfamily arXiv:1506.01961 [hep-ph]}}.

\bibitem{Agashe:2007zd}
K.~Agashe, H.~Davoudiasl, G.~Perez, and A.~Soni, ``{Warped Gravitons at the LHC
  and Beyond},'' \href{http://dx.doi.org/10.1103/PhysRevD.76.036006}{{\em Phys.
  Rev.} {\bfseries D76} (2007) 036006},
\href{http://arxiv.org/abs/hep-ph/0701186}{{\ttfamily arXiv:hep-ph/0701186
  [hep-ph]}}.

\bibitem{Aharony:2016dwx}
O.~Aharony, L.~F. Alday, A.~Bissi, and E.~Perlmutter, ``{Loops in AdS from
  Conformal Field Theory},''
  \href{http://dx.doi.org/10.1007/JHEP07(2017)036}{{\em JHEP} {\bfseries 07}
  (2017) 036},
\href{http://arxiv.org/abs/1612.03891}{{\ttfamily arXiv:1612.03891 [hep-th]}}.

\bibitem{Yuan:2017vgp}
E.~Y. Yuan, ``{Loops in the Bulk},''
\href{http://arxiv.org/abs/1710.01361}{{\ttfamily arXiv:1710.01361 [hep-th]}}.

\bibitem{Giombi:2017hpr}
S.~Giombi, C.~Sleight, and M.~Taronna, ``{Spinning AdS Loop Diagrams: Two Point
  Functions},'' \href{http://dx.doi.org/10.1007/JHEP06(2018)030}{{\em JHEP}
  {\bfseries 06} (2018) 030},
\href{http://arxiv.org/abs/1708.08404}{{\ttfamily arXiv:1708.08404 [hep-th]}}.

\bibitem{Carmi:2018qzm}
D.~Carmi, L.~Di~Pietro, and S.~Komatsu, ``{A Study of Quantum Field Theories in
  AdS at Finite Coupling},''
  \href{http://dx.doi.org/10.1007/JHEP01(2019)200}{{\em JHEP} {\bfseries 01}
  (2019) 200},
\href{http://arxiv.org/abs/1810.04185}{{\ttfamily arXiv:1810.04185 [hep-th]}}.

\bibitem{Carmi:2019ocp}
D.~Carmi, ``{Loops in AdS: From the Spectral Representation to Position
  Space},''
\href{http://arxiv.org/abs/1910.14340}{{\ttfamily arXiv:1910.14340 [hep-th]}}.

\bibitem{Meltzer:2019nbs}
D.~Meltzer, E.~Perlmutter, and A.~Sivaramakrishnan, ``{Unitarity Methods in
  AdS/CFT},''
\href{http://arxiv.org/abs/1912.09521}{{\ttfamily arXiv:1912.09521 [hep-th]}}.

\bibitem{Bzowski:2013sza}
A.~Bzowski, P.~McFadden, and K.~Skenderis, ``{Implications of conformal
  invariance in momentum space},''
  \href{http://dx.doi.org/10.1007/JHEP03(2014)111}{{\em JHEP} {\bfseries 03}
  (2014) 111},
\href{http://arxiv.org/abs/1304.7760}{{\ttfamily arXiv:1304.7760 [hep-th]}}.

\bibitem{Isono:2018rrb}
H.~Isono, T.~Noumi, and G.~Shiu, ``{Momentum space approach to crossing
  symmetric CFT correlators},''
  \href{http://dx.doi.org/10.1007/JHEP07(2018)136}{{\em JHEP} {\bfseries 07}
  (2018) 136},
\href{http://arxiv.org/abs/1805.11107}{{\ttfamily arXiv:1805.11107 [hep-th]}}.

\bibitem{Albayrak:2019asr}
S.~Albayrak, C.~Chowdhury, and S.~Kharel, ``{New relation for Witten
  diagrams},'' \href{http://dx.doi.org/10.1007/JHEP10(2019)274}{{\em JHEP}
  {\bfseries 10} (2019) 274},
\href{http://arxiv.org/abs/1904.10043}{{\ttfamily arXiv:1904.10043 [hep-th]}}.

\bibitem{Albayrak:2020isk}
S.~Albayrak, C.~Chowdhury, and S.~Kharel, ``{An \'etude of momentum space
  scalar amplitudes in AdS},''
\href{http://arxiv.org/abs/2001.06777}{{\ttfamily arXiv:2001.06777 [hep-th]}}.

\bibitem{Costantino:2020vdu}
A.~Costantino and S.~Fichet, ``{Opacity from Loops in AdS},''
  \href{http://arxiv.org/abs/2011.06603}{{\ttfamily arXiv:2011.06603
  [hep-th]}}.

\bibitem{Giddings:1999qu}
S.~B. Giddings, ``{The Boundary S matrix and the AdS to CFT dictionary},''
  \href{http://dx.doi.org/10.1103/PhysRevLett.83.2707}{{\em Phys. Rev. Lett.}
  {\bfseries 83} (1999) 2707--2710},
\href{http://arxiv.org/abs/hep-th/9903048}{{\ttfamily arXiv:hep-th/9903048
  [hep-th]}}.

\bibitem{Balasubramanian:1999ri}
V.~Balasubramanian, S.~B. Giddings, and A.~E. Lawrence, ``{What do CFTs tell us
  about Anti-de Sitter space-times?},''
  \href{http://dx.doi.org/10.1088/1126-6708/1999/03/001}{{\em JHEP} {\bfseries
  03} (1999) 001},
\href{http://arxiv.org/abs/hep-th/9902052}{{\ttfamily arXiv:hep-th/9902052
  [hep-th]}}.

\bibitem{PhysRevD.98.030001}
{\bfseries Particle Data Group} Collaboration, M.~t. Tanabashi, ``Review of
  Particle Physics,'' \href{http://dx.doi.org/10.1103/PhysRevD.98.030001}{{\em
  Phys. Rev. D} {\bfseries 98} (Aug, 2018) 030001}.
  \url{https://link.aps.org/doi/10.1103/PhysRevD.98.030001}.

\bibitem{Creminelli:2001th}
P.~Creminelli, A.~Nicolis, and R.~Rattazzi, ``{Holography and the electroweak
  phase transition},''
  \href{http://dx.doi.org/10.1088/1126-6708/2002/03/051}{{\em JHEP} {\bfseries
  03} (2002) 051},
\href{http://arxiv.org/abs/hep-th/0107141}{{\ttfamily arXiv:hep-th/0107141
  [hep-th]}}.

\bibitem{Polchinski:2002jw}
J.~Polchinski and M.~J. Strassler, ``{Deep inelastic scattering and gauge /
  string duality},''
  \href{http://dx.doi.org/10.1088/1126-6708/2003/05/012}{{\em JHEP} {\bfseries
  05} (2003) 012},
\href{http://arxiv.org/abs/hep-th/0209211}{{\ttfamily arXiv:hep-th/0209211
  [hep-th]}}.

\bibitem{Witten:1979kh}
E.~Witten, ``{Baryons in the 1/n Expansion},''
\href{http://dx.doi.org/10.1016/0550-3213(79)90232-3}{{\em Nucl. Phys.}
  {\bfseries B160} (1979) 57--115}.

\bibitem{Nastase:2007kj}
H.~Nastase, ``{Introduction to AdS-CFT},''
\href{http://arxiv.org/abs/0712.0689}{{\ttfamily arXiv:0712.0689 [hep-th]}}.

\end{thebibliography}\endgroup


\providecommand{\href}[2]{#2}\begingroup\raggedright\endgroup

\end{document}